\documentclass[11pt]{article}
\usepackage{amsmath}
\usepackage{amssymb}
\usepackage{amsthm,amsxtra}
\usepackage{epsf}
\usepackage{graphics,graphicx}
\newlength{\mytopmargin}
\newlength{\myleftmargin}
\newtheorem{lemma}{Lemma}
\newtheorem{prop}[lemma]{Proposition}
\newcommand{\ai}{{\rm Ai}}
\newcommand{\aip}{{\rm Ai}\,'}

\begin{document}
\vspace{4cm}
\noindent
{\bf Asymptotic form of the density profile for Gaussian and Laguerre
random matrix ensembles with orthogonal and symplectic symmetry}

\vspace{5mm}
\noindent
P.J. Forrester${}^{*}$, N.E. Frankel${}^\dagger$ and T.M. Garoni${}^{\dagger\dagger}$

{\small 
\noindent
${}^*$ Department of Mathematics and Statistics,
University of Melbourne, Victoria 3010, Australia ; \\
${}^\dagger$
School of Physics, University of Melbourne, Victoria 3010, Australia ; \\
${}^{\dagger\dagger}$ Institute for Mathematics and its Applications, University
of Minnesota, Minneapolis, MN 55455-0436, USA

\begin{quote}
In a recent study we have obtained correction terms to the large $N$ asymptotic
expansions of the eigenvalue density for the Gaussian unitary and Laguerre
unitary ensembles of random $N \times N$ matrices, both in the bulk and at the
soft edge of the spectrum. In the present study these results are used to
similarly analyze the eigenvalue density for Gaussian and Laguerre random matrix
ensembles with orthogonal and symplectic symmetry. As in the case of unitary
symmetry, a matching is exhibited between the asymptotic expansion of the
bulk density, expanded about the edge, and the asymptotic expansion of the
edge density, expanded into the bulk. In addition, aspects of the asymptotic
expansion of the smoothed density, which involves delta functions at the
endpoints of the support, are interpreted microscopically.

\end{quote}

}

\section{Introduction}
Perhaps the best known result in random matrix theory is the Wigner semi-circle law 
(see e.g.~\cite{Po65}).
Consider a real symmetric matrix, with elements on the diagonal i.i.d.~random variables having 
finite variance and similarly the elements above the diagonal. The Wigner semi-circle law tells
us that after appropriate scaling, the limiting eigenvalue density is given by the semi-circle
functional form
\begin{equation}\label{1}
\rho_{\rm W}(\lambda) = \left \{ \begin{array}{ll} {2 \over \pi} (1 - \lambda^2)^{1/2},
& |\lambda| < 1 \\ 0, &  |\lambda| \ge 1. \end{array} \right.
\end{equation}
As a concrete example, the Gaussian orthogonal ensemble (GOE) of real symmetric matrices is
specified by its diagonal entries being distributed according to the normal distribution
N$[0,1]$ and its upper triangular entries according to N$[0,1/\sqrt{2}]$. 
Let $\rho^{(N)}(\lambda)$ denote the eigenvalue density for  $N \times N$ 
matrices from the GOE.
After the scaling
$\sqrt{2N} \rho^{(N)}(\sqrt{2N} \lambda) \mapsto N \rho(\lambda)$ the 
$N \to \infty$ limiting form of $\rho^{(N)}(\lambda)$  
is given by (\ref{1}).

The functional form (\ref{1}) has implications with respect to averaging a so called linear
statistic $A = \sum_{j=1}^N a(\lambda_j)$ over the spectrum of random real symmetric matrices.
Thus, if the $N \to \infty$
scaling is such that ${\alpha_N \over N} \rho^{(N)}(\alpha_N \lambda) \to
\rho_{\rm W}(\lambda)$, $a(\alpha_N \lambda) \to \tilde{a}(\lambda)$ for some
$\alpha_N$ then
\begin{equation}\label{2}
\langle A \rangle := \int_{-\infty}^\infty \rho^{(N)}(\lambda) a(\lambda) \, d \lambda
\: \sim \: N \int_{-1}^1 \rho_{\rm W}(\lambda)  \tilde{a}(\lambda) \, d \lambda.
\end{equation}
The result (\ref{2}) immediately draws attention to corrections to the Wigner semi-circle law.
Indeed in studies of the full distribution of linear
statistics averaged over the GOE and certain of its generalizations, it is necessary to compute
the O$(1)$ term in the asymptotic expansion of (\ref{2}) \cite{Jo98}. For this one seeks
the asymptotic
expansion of $\bar{\rho}^{(N)}(\lambda)$, where $\bar{\rho}^{(N)}(\lambda)$ is the signed
measure (smoothed density) such that
$$
\int_{-\infty}^\infty \rho^{(N)}(\lambda) a(\lambda) \, d \lambda =
\int_{-\infty}^\infty \bar{\rho}^{(N)}(\lambda) a(\lambda) \, d \lambda 
$$
to all orders in the corresponding asymptotic expansions.
In the case of the
GOE itself the O$(1)$ term is known \cite{VZ84,DJ90,Jo98}, and one has
\begin{equation}\label{3}
{\sqrt{2N} \over N}
 \bar{\rho}^{(N)}(\sqrt{2N} \lambda) \: \sim \: \rho_{\rm W}(\lambda) +
{1 \over N} \bigg ( {1 \over 4} \Big ( \delta(\lambda - 1) + \delta (\lambda + 1) 
\Big ) -  {1 \over 2 \pi} {1 \over \sqrt{1 - \lambda^2} }
\chi_{|\lambda| < 1} \bigg )
\end{equation}
where $\chi_T = 1$ if $T$ is true and $\chi_T = 0$ otherwise. 

The correction term in (\ref{3}) exhibits a most remarkable  feature, namely delta functions at the
edge of the support of the spectrum. The appearance of the delta functions at a microscopic
level, when one seeks directly the asymptotic expansion of ${\sqrt{2N} \over N}
 \rho^{(N)}(\sqrt{2N} \lambda)$ rather than the asymptotics of the smoothed quantity
$ \bar{\rho}^{(N)}(\lambda)$ 
has not, to the best of our knowledge, been previously studied. One of the purposes of
this paper is to undertake such a study for the 
Gaussian and Laguerre ensembles in
random matrix theory. Each of the three symmetry classes, 
orthogonal ($\beta = 1$), unitary ($\beta = 2$) and symplectic ($\beta = 4$)
will be considered. For the Gaussian ensemble 
it is known  \cite{Jo98} that (\ref{3}) then generalizes to read
\begin{equation}\label{1.3'}
{\sqrt{2N} \over N}
 \bar{\rho}^{(N)}(\sqrt{2N} \lambda) \: \sim \: \rho_{\rm W}(\lambda) +
{1 \over N} \Big ( {1 \over \beta} - {1 \over 2} \Big )
\bigg ( {1 \over 2} \Big ( \delta(\lambda - 1) + \delta (\lambda + 1)
\Big ) -  {1 \over  \pi} {1 \over \sqrt{1 - \lambda^2} }
\chi_{|\lambda| < 1} \bigg ).
\end{equation}
Our task is to relate this expansion to the asymptotic expansion of the density
itself.

The expansion (\ref{1.3'}) clearly shows both a bulk effect and an edge effect.
This is in keeping with there being both a (global) bulk regime, and an edge regime
which must be treated separately in the asymptotic analysis. As these expansions
relate to the same quantity, one would expect there to be a matching in an
appropriate limit. This topic, initiated in \cite{GFF05} for the GUE and LUE, is
another main theme of the present work.

We begin in Section 2 by recalling the results from  \cite{GFF05} relating to the asymptotic
expansions of the global bulk density, and the soft edge density, in the GUE and LUE. We then
proceed to write down higher order terms in these asymptotic expansions (obtained from the method
of \cite{GFF05}). These higher order terms are then used to further investigate the matching
phenomenon alluded to above.

In Section 3 formulas required in the study of the asymptotics of the density in the Gaussian
and Laguerre ensembles with orthogonal and symplectic symmetry are gathered. These formulas are
used in Section 4 to study the corresponding global density asymptotic expansions, and in
Section 5 to study the soft edge density asymptotic expansions. In Section 6 we use the
results of Section 2, 4 and 5 to study our main topics of interest, namely the matching
between the bulk and edge asymptotic expansions, and the microscopic origin of the delta
functions in (\ref{1.3'}) and its Laguerre analogue.

\section{The Gaussian and Laguerre ensembles with unitary symmetry}
\setcounter{equation}{0}
\subsection{Definitions and summary of known results}
The Gaussian unitary ensemble consists of the set of Hermitian matrices with diagonal
entries distributed according to the normal distribution N$[0,1/\sqrt{2}]$ and with upper
triangular entries distributed
according to N$[0,1/2] + i {\rm N}[0,1/2]$. The corresponding eigenvalue
p.d.f.~is given by
\begin{equation}\label{2.1}
{1 \over C} \prod_{l=1}^N e^{-x_l^2} \prod_{1 \le j < k \le N} (x_k - x_j)^2
\end{equation}
where here and below $C$ denotes {\it some} normalization constant.

The Laguerre unitary ensemble can be specified by matrices $X = A^\dagger A$ where $A$ is a
$n \times m$ $(n \ge m)$ complex Gaussian matrix with entries distributed according to
N$[0,1/\sqrt{2}] + i {\rm N}[0,1/\sqrt{2}]$. All the eigenvalues are non-negative and have the
joint distribution
\begin{equation}\label{2.2}
{1 \over C} \prod_{l=1}^m x_l^\alpha e^{-x_l} \prod_{1 \le j < k \le m} (x_k - x_j)^2
\end{equation}
where $\alpha = n - m$.

The eigenvalue p.d.f.s (\ref{2.1}) and (\ref{2.2}) are special cases of the functional form
\begin{equation}\label{Ug}
{\rm UE}_N(g_2) : = {1 \over C} \prod_{l=1}^N g_2(x_l) \prod_{1 \le j < k \le N}
(x_k - x_j)^2
\end{equation}
defining a general matrix ensemble with unitary symmetry in terms of its eigenvalue
p.d.f. Thus
$$
{\rm Ev}({\rm GUE}_N) = {\rm UE}_N(e^{-x^2}), \qquad
{\rm Ev}({\rm LUE}_N) = {\rm UE}_N(x^\alpha e^{-x})
$$
where Ev$(M)$ denotes the eigenvalue p.d.f.~of the matrix ensemble $M$.

It is a basic result in random matrix theory (see e.g.~\cite{Fo02}) that the eigenvalue
density for the ensemble (\ref{Ug}) can be written in terms of the monic polynomials
$\{p_n(x)\}$ orthogonal with respect to the weight $g_2(x)$. Thus with $(p_n,p_n)_2$ denoting
the corresponding normalizations we have
\begin{equation}\label{Uga}
\rho(x;{\rm UE}_N(g_2)) = g_2(x) \sum_{j=0}^{N-1}
{(p_j(x))^2 \over (p_j,p_j)_2}.
\end{equation}

In a recent study Kalisch and Braak \cite{KB02} have obtained the leading correction term to the
Wigner semi-circle law for the asymptotic expansion of (\ref{Uga}) in the case of the GUE.

\begin{prop}
Let $-1 < X < 1$ be fixed. One has
\begin{equation}\label{2.3}
{1 \over N} \rho( X; {\rm UE}_N(e^{-2 N x^2})) \: \sim \:
\rho_{\rm W}(X) - {2 \cos (2 N \pi P_{\rm W}(X) ) \over \pi^3 \rho_{\rm W}^2(X) }
{1 \over N} + {\rm O} \Big ( {1 \over N^2} \Big ),
\end{equation}
where $\rho_{\rm W}(x)$ is given by (\ref{1}) and
\begin{equation}\label{PW}
P_{\rm W}(x) = 1 + {x \over 2} \rho_{\rm W}(x) - {1 \over \pi} {\rm Arccos} \, x.
\end{equation}
\end{prop}

The methods in \cite{KB02} are particular to the 
Gaussian ensembles, relying on an integral
formula coming from the supersymmetry method. Subsequently the present authors
\cite{GFF05} have introduced a different strategy which reclaims (\ref{2.3}),
and furthermore applies equally well to the Laguerre case, for which the following result was
obtained.

\begin{prop}
Let $0 < X < 1$ be fixed. We have
\begin{eqnarray}\label{2.7}
&&
{1 \over N}  \rho(X; {\rm UE}_N(x^\alpha e^{-4 N x})) \: \sim \: \rho_{\rm MP}(X)
\nonumber \\
&& \qquad
 - \Big (
{\cos ((2N+\alpha) \pi P_{\rm MP}(X) - 
\alpha \pi (1 + X \rho_{\rm MP}(X)) \over \pi^3 X^2
\rho_{\rm MP}^2(X) } -
{\alpha  \over \pi^2 X \rho_{\rm MP}(X) } \Big ) {1 \over N} +
O\Big ( {1 \over N^2} \Big )
\end{eqnarray}
where
\begin{equation}
\rho_{\rm MP}(x)  :=  {2 \over \pi} \sqrt{{1 \over x} - 1}, \qquad 
P_{\rm MP}(x) = 1 + x \rho_{\rm MP}(x) - {2 \over \pi} {\rm Arccos} \sqrt{x}
\end{equation}
(the subscript MP denotes Mar\v{c}enko-Pastur, who first derived the limit law giving the leading term
in this expression).
\end{prop}

The strategy of \cite{GFF05} was to use integral representations of the product of orthogonal
polynomials which result from the Christoffel-Darboux summation of (\ref{Uga}). In addition to
yielding the asymptotics of the global bulk density, it also gave asymptotics of the
density at the so called soft edge. This is the name given to the boundary of the
support at leading order, with the feature that the density, appropriately scaled, is non-zero on both sides
of this boundary.

\begin{prop}
Let $\xi$ be fixed. 
For the GUE
\begin{eqnarray}\label{GUE edge result}
&& \frac{1}{2 N^{2/3}} \rho\left(1+\frac{\xi}{2 N^{2/3}};{\rm UE}_N(e^{-2N x^2})\right)
=
[\aip(\xi)]^2-\xi [\ai(\xi)]^2
\nonumber \\&& \qquad
-\frac{1}{20}\left(3\xi^2[\ai(\xi)]^2 -2\xi[\aip(\xi)]^2 -3\ai(\xi)\aip(\xi)\right)\frac{1}{N^{2/3}}
+O\left(\frac{1}{N}\right),
\end{eqnarray}
while for the LUE
\begin{eqnarray}\label{LUE edge result}
&& \frac{1}{(2N)^{2/3}}\rho\left(1+\frac{\xi}{(2N)^{2/3}};{\rm UE}_N(x^{\alpha}
e^{-4Nx})
\right)
\nonumber \\ && \quad =
([\aip(\xi)]^2-\xi[\ai(\xi)]^2)
+ \frac{\alpha}{2^{1/3}} [\ai(\xi)]^2  \frac{1}{N^{1/3}}
\nonumber \\&& \qquad
+\!\frac{2^{1/3}}{10}\!\left(3\xi^2[\ai(\xi)]^2-2\xi[\aip(\xi)]^2+(2-5\alpha^2)\ai(\xi)\aip(\xi) \right)\frac{1}{N^{2/3}}
+O\left(\frac{1}{N}\right).
\end{eqnarray}
\end{prop}

\subsection{Matching of the bulk and edge expansions}\label{s2.2}
We pursue the matching phenomenon observed in 
\cite{GFF05} between the asymptotic expansion of the bulk density, expanded about the
soft edge, and the asymptotic expansion of the edge density, expanded into the bulk.
Explicitly, it was found that setting
\begin{equation}\label{XZ}
X = 1 + \xi/2 N^{2/3}, \qquad (\xi < 0 )
\end{equation}
in the asymptotic expansion (\ref{2.3}) multiplied by $N^{2/3}/2$, and then expanding in
$N$, a matching is obtained with
the $\xi \to - \infty$ asymptotic expansion of the right hand side of
(\ref{GUE edge result}). This was checked on terms in $\xi$ in the latter accessible
by expanding the former; on this point we note that terms of all orders in
inverse powers of $N$ in
(\ref{2.3}), will after substitution of (\ref{XZ}) contribute to each term in the
expansion of
(\ref{GUE edge result}). A similar matching was observed between the bulk LUE density
expanded with the substitution (\ref{XZ}), and the $\xi \to - \infty$ 
asymptotic expansion of the $O(1)$ and $O(N^{-1/3})$ terms of the edge density
(\ref{LUE edge result}).  

Based on this evidence, the hypothesis was put forward in \cite{GFF05} that the
matching persists between all terms in $\xi$, and at all orders in inverse 
(fractional) powers of $N$. Here we probe this hypothesis further by
extending the asymptotic expansions (\ref{2.3}),  (\ref{2.7}),
(\ref{GUE edge result}) and (\ref{LUE edge result}).
As explained in \cite{GFF05}, the orthogonal polynomial method readily allows for the
computation of higher order terms, which we compute to be given as follows.

\begin{prop}
The $O(1/N^2)$ term in (\ref{2.3}) is
\begin{equation}\label{2.5a}
\bigg ( {1 \over 16 \pi (1 - X^2)^{5/2} } +
{X (15 + 2X^2) \sin [ 2 N \pi P_{\rm W}(X) ] \over 48 \pi (1 - X^2)^{5/2} } \bigg )
{1 \over N^2};
\end{equation} 
the $O(1/N^3)$ term in (\ref{2.3}) is
\begin{equation}\label{2.5b}
{180 + 981 X^2 + 60 X^4 + 4 X^6  \over 2304 \pi (1 - X^2)^{4} }
 \cos [ 2 N \pi P_{\rm W} (X) ] {1 \over N^3};
\end{equation}
the oscillatory $O(1/N^4)$ term in (\ref{2.3}) is
\begin{equation}\label{2.5c}
-X {323190  + 647055 X^2 + 20358  X^4 + 6084 X^6 - 1112 X^8 \over 829440 \pi (1 - X^2)^{11/2} }
 \sin [ 2 N \pi P_{\rm W} (X) ] {1 \over N^4};
\end{equation}
the $O(1/N^2)$ term in (\ref{2.7}) is
\begin{eqnarray}\label{2.7a}
&&
\bigg ( \Big (  { \alpha \over 8 \pi (1 - X)^{2} } \Big )
\cos [ 2N \pi P_{\rm MP}(X) - 2 \alpha {\rm Arc cos} \, \sqrt{X} ] 
\nonumber \\
&& +
\Big ( {-3 + 12 X + 8 X^2 + 12 (-1 + X) (-1 + 2 X) \alpha^2 \over 192 \pi
(1 - X)^{5/2} X^{3/2} } \Big )
\sin [ 2N \pi P_{\rm MP}(X) - 2 \alpha {\rm Arc cos} \, \sqrt{X} ]
\nonumber \\
&&
+ { 1 + 4(-1 + X) \alpha^2 \over 64 \pi (1 - X)^{5/2} X^{3/2} }
\bigg ) {1 \over N^2};
\end{eqnarray}
the $O(1/N)$ term in (\ref{GUE edge result}) is
\begin{equation}\label{GN}
\bigg ( \Big ( - {\xi \over 12} + {3 \xi^4 \over 40} + {\xi^7 \over 48} \Big )
(\ai(\xi))^2 - {3 \xi^2 \over 40} \ai(\xi) \aip(\xi)  +
 \Big (  {1 \over 12} - {\xi^3 \over 20} - {\xi^6 \over 48} \Big )
(\aip(\xi))^2 \bigg ) {1 \over N};
\end{equation}
the $O(1/N)$ term in (\ref{LUE edge result}) is
\begin{equation}
\alpha \bigg ( \Big ( - {7 \xi \over 15} + {\alpha^2  \xi \over 6}  \Big )
(\ai(\xi))^2 - { \xi^2 \over 5} \ai(\xi) \aip(\xi)  +
 \Big ( - {1 \over 6} + {\alpha^2 \over 6} \Big )
(\aip(\xi))^2 \bigg ) {1 \over N}.
\end{equation}
\end{prop}

Considering first the GUE,
we now substitute (\ref{XZ}) in (\ref{2.3}) extended by (\ref{2.5a}), (\ref{2.5b}),
(\ref{2.5c}). 
Expanding the asymptotic series (an operation we denote by $\mathop{\sim}\limits^\cdot$) gives
the new asymptotic series in $N$,
\begin{eqnarray}\label{tm}
&& {1 \over N^{2/3}} \rho\Big ( 1 + {\xi \over 2 N^{2/3} } ; {\rm UE}_N(e^{-2N x^2}) \Big ) 
\nonumber \\
&& \mathop{\sim}\limits^\cdot
\bigg ( {2 \sqrt{|\xi|} \over \pi} + {1 \over 16 \pi |\xi|^{5/2} } +
\Big ( - { 1 \over 2 \pi |\xi| } + {1225 \over 2304 \pi \xi^4 } \Big )
 \cos  (4 |\xi|^{3/2}/3) - 
{17 \sin (4 |\xi|^{3/2}/3) \over 48 \pi |\xi|^{5/2} } \bigg ) \nonumber \\
&&
\qquad + \bigg ( - { |\xi|^{3/2} \over 4 \pi} + {5 \over 128 \pi  |\xi|^{3/2} } +
\Big ( - {43 \over 480 \pi} - {23695 \over 331776 \pi |\xi|^3} \Big )  \cos  (4 |\xi|^{3/2}/3)
\nonumber \\
&&
\qquad 
+ \Big ( {233 \over 4608} - {|\xi|^3 \over 20} \Big )
{\sin (4 |\xi|^{3/2}/3) \over \pi  |\xi|^{3/2} } \bigg ) {1 \over N^{2/3} }
\nonumber \\
&& \qquad + O\Big ( {1 \over N^{4/3}} \Big ).
\end{eqnarray}
On the other hand, making use of the asymptotic series \cite{Ol74}
\begin{equation}\label{aia}
\ai(-z) \: \sim \: \pi^{-1/2} z^{-1/4} \Big ( \sin(\zeta + \pi/4) \sum_{k=0}^\infty
(-1)^k c_{2k} \zeta^{-2k} -
\cos(\zeta + \pi/4) \sum_{k=0}^\infty
(-1)^k c_{2k+1} \zeta^{-2k-1} \Big ),
\end{equation}
where
$$
\zeta = {2 \over 3} z^{3/2}, \qquad c_0 = 1, \qquad
c_k = {(2k+1)(2k+3) \cdots (6k-1) \over (216)^k k! } \: \: (k \ge 1),
$$
the $\xi \to  -\infty$ expansion of the $O(1)$ and $O(1/N^{2/3})$ terms in
(\ref{GUE edge result}) can readily be computed. Agreement is found with all terms
in (\ref{tm}) except the one involving the fraction $23695/331776$. Thus, even though no terms
$O(1/N)$ have yet appeared, whereas (\ref{GN}) has a term at this order, the evidence is
still in favor of a matching between all terms in $\xi$, and at all orders in inverse powers
of $N$. However this matching cannot be fully exhibited at any order in $N$ in the expansion
of (\ref{GUE edge result}) without knowledge of all terms in the asymptotic expansion of
(\ref{2.3}).

As already mentioned,
a similar matching phenomenon was observed in \cite{GFF05} in the case of the LUE, and
conjectured to hold at general orders as for the GUE. Further evidence for this conjecture
can be obtained by substituting
(\ref{XZ}) in (\ref{2.7}) extended by (\ref{2.7a}), expanding as a series in inverse powers of
$N$, and comparing against the $\xi \to -\infty$ expansion of (\ref{LUE edge result}). The former
operation gives
\begin{eqnarray}
\lefteqn{
{1 \over N^{2/3}} \rho \Big ( 1 + {\xi \over 2 N^{2/3}} ;
{\rm UE}_N(x^\alpha e^{-4N x}) \Big ) } \nonumber \\
&&
  \mathop{\sim}\limits^\cdot 
\Big ( {2^{2/3} |\xi|^{1/2} \over \pi} -
{\cos(4 |\xi|^{3/2}/3) \over 2^{4/3} \pi |\xi| } \Big )
+ {\alpha (1 + \sin(4|\xi|^{3/2}/3)) \over 2^{2/3} \pi |\xi|^{1/2} N^{1/3}} \nonumber \\
&& \quad +
{1 \over 160 \pi |\xi|^{3/2} N^{2/3} } \Big (
5 - 20 \alpha^2 + 80|\xi|^3 + 40(-1+2 \alpha^2)  |\xi|^{3/2} \cos(4|\xi|^{3/2}/3) 
 \nonumber \\ && \quad +
(5 - 20\alpha^2 + 16 |\xi|^3) \sin(4|\xi|^{3/2}/3) \Big ) + O \Big ( {1 \over N} \Big ).
\end{eqnarray}
The latter operation gives agreement with this expansion at $O(1)$, $O(1/N^{1/3})$ and for
the terms involving factors of $|\xi|^3$ at $O(1/N^{2/3})$. This is consistent with the
matching hypothesis.

\section{The Gaussian and Laguerre ensembles with orthogonal and symplectic symmetry --
general formulas}
\setcounter{equation}{0}
The Gaussian orthogonal ensemble has already been defined in the Introduction. At the level of an
eigenvalue p.d.f., the GOE can be defined by the joint distribution
$$
{1 \over C} \prod_{l=1}^N e^{-x_l^2/2} \prod_{1 \le j < k \le N} |x_k - x_j|.
$$
Likewise the Gaussian symplectic ensemble, Laguerre orthogonal ensemble and  Laguerre symplectic
ensemble can be specified either in terms of the distribution of certain classes of random
matrices (real symmetric matrices in the cases of orthogonal symmetry, and Hermitian matrices
with real quaternion elements in the cases of symplectic symmetry), or in terms of the
functional form of the eigenvalue p.d.f.~(see e.g.~\cite{Fo02}). Here we will note only the latter,
which in the case of the GSE reads
$$
{1 \over C} \prod_{l=1}^N e^{- 2 x_l^2} \prod_{1 \le j < k \le N} (x_k - x_j)^4;
$$
for the LOE reads
$$
{1 \over C} \prod_{l=1}^N x_l^{\alpha/2} e^{-x_l/2} \prod_{1 \le j < k \le N}
|x_k - x_j|;
$$
and for the LSE reads
$$
{1 \over C} \prod_{l=1}^N x_l^{ 2\alpha} e^{-2 x_l} \prod_{1 \le j < k \le N}
(x_k - x_j)^4.
$$
For the Laguerre ensembles one requires the eigenvalues be positive and thus $x_l > 0$
$(l=1,\dots,N)$. 
Thus we see that if we define a matrix ensemble with orthogonal and symplectic symmetry by the
eigenvalue p.d.f.s
$$
{\rm OE}_N(g_1) := {1 \over C} \prod_{l=1}^N g_1(x_l) \prod_{1 \le j < k \le N} |x_k - x_j|
$$
and
$$
{\rm SE}_N(g_4) :=  {1 \over C} \prod_{l=1}^N g_4(x_l) \prod_{1 \le j < k \le N} (x_k - x_j)^4
$$
respectively, then we have
\begin{equation}
\begin{array}{ll} {\rm Ev}({\rm GOE}_N) = {\rm OE}_N(e^{-x^2/2}), &
{\rm Ev}({\rm LOE}_N) = {\rm OE}_N(x^{\alpha/2} e^{-x/2}) \\
{\rm Ev}({\rm GSE}_N) = {\rm SE}_N(e^{-2 x^2}), &
{\rm Ev}({\rm LSE}_N) = {\rm SE}_N(x^{2 \alpha} e^{-x}). \end{array}
\end{equation}

In the work \cite{AFNV00}, the eigenvalue densities for the ensembles OE$(g_1)$ and SE$(g_4)$ were
computed for all the so called classical weights
\begin{equation}\label{7.86'}
g_1(x) = \left \{ \begin{array}{ll}
e^{-x^2/2}, & {\rm Gaussian} \\
x^{(\alpha -1)/2} e^{-x/2} \: (x > 0), & {\rm Laguerre} \\
(1-x)^{(a-1)/2}(1+x)^{(b-1)/2} \: (-1 < x < 1), & {\rm Jacobi} \\
(1+x^2)^{-(\alpha + 1)/2}, & {\rm Cauchy}
\end{array} \right.
\end{equation}
\begin{equation}\label{7.48'}
g_4(x) =
\left \{ \begin{array}{ll} e^{-x^2}, & {\rm Hermite} \\
x^{\alpha +1} e^{-x}, & {\rm Laguerre} \\
(1 - x)^{a+1} (1 + x)^{b+1}, & {\rm Jacobi} \\
(1+x^2)^{-(\alpha - 1)}, & {\rm Cauchy}
\end{array} \right.
\end{equation}
in terms of a formula depending on the symmetry class only. Thus for ensembles
OE${}_N(g_1)$ with classical weights
(\ref{7.86'}) one has
\begin{eqnarray}\label{7.4.sum1}
\lefteqn{\rho(x;{\rm OE}_N(g_1(x)))   =  
\rho(x;{\rm UE}_{N-1}(g_2(x))) } 
\nonumber \\ &&  +  {c_{N-2} \over (p_{N-2}, p_{N-2})_2 
(p_{N-1}, p_{N-1})_2 } g_1(x) p_{N-1}(x)
{1 \over 2} \int_{-\infty}^\infty {\rm sgn}(x-t) p_{N-2}(t)
g_1(t) \, dt,
\end{eqnarray}
while for ensembles SE${}_N(g_4)$ with classical weights (\ref{7.48'}) one has
\begin{eqnarray}\label{7.nfs}
\rho(x;{\rm SE}_N(g_4(x)))
& = &
{1 \over 2} \rho(x;{\rm UE}_{2N}(g_2))
\nonumber \\ &&
- g_1(x)  {c_{2N-1} p_{2N}(x) \over
2 (p_{2N}, p_{2N})_2 (p_{2N-1}, p_{2N-1})_2 }
\int_x^\infty  g_1(t)  
p_{2N-1}(t) \, dt.
\end{eqnarray}

In (\ref{7.4.sum1}) and (\ref{7.nfs}),
\begin{equation}\label{7.48}
g_2(x) =
\left \{ \begin{array}{ll} e^{-x^2}, & {\rm Hermite} \\
x^{\alpha } e^{-x}, & {\rm Laguerre} \\
(1 - x)^{a} (1 + x)^{b}, & {\rm Jacobi} \\
(1+x^2)^{-\alpha }, & {\rm Cauchy}
\end{array} \right.
\end{equation}
while
\begin{equation}\label{7.48a}
{c_j \over  (p_{j}, p_{j})_2 }
=  \left \{ \begin{array}{ll} 1, & {\rm Hermite} \\
{1 \over 2} , & {\rm Laguerre}.
\end{array} \right. 
\end{equation}
The quantities $\{p_n(x)\}$ are the monic classical orthogonal polynomials with respect
to the weights (\ref{7.48}), and $(p_n,p_n)_2$ the corresponding normalizations. Thus
in the Gaussian case
\begin{equation}\label{G1}
p_n(x) =  2^{-n}  H_n(x), \qquad (p_n,p_n)_2  =  \pi^{1/2}
2^{-n}  n!
\end{equation}
while in the Laguerre case
\begin{equation}\label{L1}
p_n(x)  =  (-1)^n n! L_n^\alpha(x), \qquad 
(p_n,p_n)_2  =  \Gamma(n+1) \Gamma (\alpha + n + 1) . 
\end{equation} 

Essential tools in our subsequent analysis of the asymptotic forms of
(\ref{7.4.sum1}) and (\ref{7.nfs}) are particular asymptotic formulas for the Hermite
and Laguerre polynomials. 
Consider first the bulk region.
In the case of the Hermite polynomials, the formula is due to
Plancherel and Rotach \cite{PR34}. It tells us that with
$$
x = (2n+1)^{1/2} \cos \phi, \qquad \epsilon \le \phi \le \pi - \epsilon,
$$
we have
$$
e^{-x^2/2} H_n(x) = 2^{n/2 + 1/4} (n!)^{1/2} (\pi n)^{-1/4} (\sin \phi)^{-1/2}
\Big ( \sin \big ( (n/2 + 1/4)(\sin 2 \phi - 2 \phi) + 3 \pi/4 ) + O(n^{-1})
\Big ).
$$
Setting
$$
\sqrt{2N} X = (2(N+m)+1)^{1/2} \cos \phi 
$$
with $-1 < X < 1$ fixed we deduce from this that for $m$ fixed
\begin{equation}\label{Hm}
H_{N+m}(\sqrt{2N} X) = \Big ( {2 \over \pi} \Big )^{1/4}
{2^{m/2 + N/2} \over (1 - X^2)^{1/4} } N^{m/2 - 1/4} (N!)^{1/2} e^{N X^2} g_{m,N}^{(H)}(X)
\Big ( 1 + O\Big ({1 \over N} \Big ) \Big )
\end{equation}
where
\begin{equation}\label{3.11}
g_{m,N}^{(H)}(x) := \cos \Big ( N x\sqrt{1-x^2} + (N+1/2) {\rm Arcsin} \, x - N \pi/2 - m
{\rm Arccos} \, x \Big ).
\end{equation} 

The Plancherel-Rotach formula (\ref{Hm}) was extended by Moecklin to the Laguerre polynomials
\cite{Moecklin34}. With
$$
x = (4n + 2 \alpha + 2) \cos^2 \phi, \qquad \epsilon \le \phi \le \pi/2 - \epsilon n^{-1/2}
$$
it reads
\begin{eqnarray*}
e^{-x/2} L_n^{(\alpha)}(x) & = & (-1)^n (\pi \sin \phi)^{-1/2} x^{-\alpha/2 - 1/4} 
n^{\alpha/2 - 1/4} \\
&& \times \Big ( \sin \Big ( (n + (\alpha + 1)/2) (\sin 2 \phi - 2 \phi) + 3 \pi / 4
\Big ) + (nx)^{-1/2} O(1) \Big ).
\end{eqnarray*}
Setting 
$$
4 n X = (4(n+m) + 2\alpha + 2) \cos^2 \phi
$$
with $m$ fixed and $\epsilon/n < X < 1$ we deduce from this that
\begin{equation}\label{Lm}
x^{\alpha / 2} e^{-x/2} L_{n+m}^{(\alpha)}(x) 
\Big |_{x = 4n X}  =  (-1)^{n+m} (2 \pi \sqrt{X(1-X)} )^{-1/2} n^{\alpha/2 - 1/2}
\Big ( g_{m,n}^{(L)}(X) + O({1 \over n}) \Big )
\end{equation}
where
\begin{equation}\label{gL}
g_{m,n}^{(L)}(X) := \sin \Big ( 2n( \sqrt{X(1-X)} - {\rm Arccos} \, \sqrt{X}) -
(2m+\alpha + 1)  {\rm Arccos} \, \sqrt{X} + 3 \pi / 4 \Big ).
\end{equation}

We turn our attention now to the (soft) spectrum edge. In the Hermite case, the formula of
Plancherel and Rotach tells us that with
\begin{equation}\label{xt}
x = (2N)^{1/2} + 2^{-1/2} N^{-1/6} t
\end{equation}
we have
$$
\exp(-x^2/2) H_N(x) = \pi^{1/4} 2^{N/2 + 1/4} (N!)^{1/2} N^{-1/12}
\Big \{  \ai(t) + O(N^{-2/3}) \Big \},
$$
where $\ai(t)$ denotes the Airy function.
It follows from this that with $x$ again related to $t$ by (\ref{xt})
\begin{equation}\label{xt1}
\exp(-x^2/2) H_{N+m}(x) = (2N)^{m/2} \pi^{1/4}  2^{N/2 + 1/4} (N!)^{1/2} N^{-1/12}
\Big \{ \ai(t) - {m \over N^{1/3} } \aip(t) + O(N^{-2/3}) \Big \}.
\end{equation} 
In the Laguerre case, Szeg\"o \cite{Sz75} gives that with 
\begin{equation}\label{xta}
x = 4N + 2 \alpha + 2 + 2 (2N)^{1/3} t
\end{equation}
we have
$$
e^{-x/2} L_N^\alpha(x) = (-1)^N 2^{-\alpha - 1/3} N^{-1/3} \Big ( \ai(t) + O(N^{-2/3}) \Big ).
$$
It then follows that for fixed $p$
\begin{equation}\label{3.17} 
e^{-x/2} L_{N+p}^\alpha(x) \Big |_{x = 4N + 2(2N)^{1/3} \xi}
= (-1)^{N+p}
2^{-\alpha - 1/3} N^{-1/3} \Big ( \ai(\xi) - {{2p + \alpha + 1}
\over (2N)^{1/3} } \aip(\xi)+ O(N^{-2/3}) \Big ).
\end{equation}

\section{Asymptotic expansions in the bulk}
\subsection{GOE and GSE in the bulk}
\setcounter{equation}{0}
As with the GUE, the asymptotic expansion of the bulk density for the GOE and GSE has been
carried out by Kalisch and Braak \cite{KB02}. But as their method is particular to the Gaussian
case, we give an alternative method which can be extended to the Laguerre case.

Consider first the GOE. Substituting the appropriate formula from (\ref{7.48a}), and (\ref{G1}),
in (\ref{7.4.sum1}) we see after minor manipulation that
\begin{equation}\label{H1}
\rho(x;{\rm OE}_N(e^{-x^2/2}))   = 
\rho(x;{\rm UE}_{N-1}(e^{-x^2})) + {e^{-x^2/2} H_{N-1}(x) \over 2^{N-1} \pi^{1/2} (N-2)! }
\int_0^x e^{-t^2/2} H_{N-2}(t) \, dt.
\end{equation}
Also, making use of (\ref{Uga}) shows
\begin{equation}\label{H1'}
\rho(x;{\rm UE}_{N-1}(e^{-x^2})) = \rho(x;{\rm UE}_{N}(e^{-x^2})) -
{2^{-(N-1)} e^{-x^2} \over \pi^{1/2} (N-1)! } ( H_{N-1}(x) )^2.
\end{equation}
From (\ref{H1}) and (\ref{H1'}) the following asymptotic formula for the bulk eigenvalue
density is obtained.

\begin{prop}\label{p1a}
Let $-1 < X < 1$ be fixed. We have
\begin{equation}\label{H1a}
{1 \over N} \rho(X;{\rm OE}_N(e^{-N x^2}))  \: \sim \: 
{2 \over \pi} \sqrt{1 - X^2} - {1 \over 2 \pi N \sqrt{1 - X^2} } +
O \Big ({1 \over N^2} \Big ).
\end{equation}
\end{prop}

\noindent
Proof. \quad 
First we note that
\begin{equation}\label{3.ke}
\rho(X;{\rm OE}_N(e^{-Nx^2})) = \sqrt{2N} \rho(\sqrt{2N} X; {\rm OE}_N(e^{-x^2/2})),
\end{equation}
and we proceed to analyze the large $N$ form of the right hand side using (\ref{H1}). In
relation to the latter, by a
simple change of variables,
$$
\int_0^{\sqrt{2N} X} e^{-t^2/2} H_{N-2}(t) \, dt =
\sqrt{2N} \int_0^X  e^{-t^2/2} H_{N-2}(t)  \Big |_{t = \sqrt{2N} T} \, dT,
$$
while
making use of (\ref{Hm}) shows
\begin{eqnarray}\label{HTa}
\lefteqn{
\sqrt{2N} \int_0^X e^{-t^2/2} H_{N-2}(t) \Big |_{t = \sqrt{2N} T} \, dT }
\nonumber \\
&&
\sim  \sqrt{2N} 
\Big ( {2 \over \pi} \Big )^{1/4} {2^{-1 + N/2} \over (1 - T^2)^{1/4} }
N^{-5/4} (N!)^{1/2} \int_0^X {g_{-2,N}^{(H)}(T) \over 
(1 - T^2)^{1/4} } \, dT.
\end{eqnarray}
The leading contribution to the integral for large $N$ comes from the neighborhood of the endpoints
$T=0$ and $T=X$. About $T=0$
$$
g_{m,N}^{(H)}(T) \: \sim \: \cos \Big ( 2N T - (N+m) \pi/2 + O(T^2) \Big ),
$$
while about $T=X$
\begin{eqnarray*}
g_{m,N}^{(H)}(T) & \sim &  \cos \Big ( N X \sqrt{1 - X^2} + (N+1/2) {\rm Arcsin} \, X  \\
&& \qquad - N \pi /2 -
m {\rm Arccos} \, X + 2N \sqrt{1 - X^2} (T-X) +
O((T-X)^2)\Big ).
\end{eqnarray*}
Thus we have
\begin{equation}\label{HTb}
\int_0^X {g_{-2,N}^{(H)}(T) \over 
(1 - T^2)^{1/4} } \, dT \: \sim \: {1 \over 2N  (1 - X^2)^{3/4} }
\tilde{g}_{-2,N}^{(H)}(X)
\end{equation}
where
\begin{equation}\label{3.18a}
\tilde{g}_{m,N}^{(H)}(x) := \sin \Big ( N x \sqrt{1-x^2} + (N + 1/2) {\rm Arcsin} \, x -
N \pi / 2 -
m {\rm Arccos} \, x \Big )
\end{equation}
and use has been made of the fact that $N$ is assumed even in (\ref{7.4.sum1}).

We read off from (\ref{Hm}) that
\begin{equation}\label{H3}
e^{-x^2} H_{N-1}(x) \Big |_{x = \sqrt{2N} X} =
\Big ( {2 \over \pi} \Big )^{1/4} {2^{-1/2 + N/2} \over (1 - X^2)^{1/4} }
N^{-3/4} (N!)^{1/2} g_{-1,N}^{(H)}(X) \Big ( 1 + O\Big ({1 \over N} \Big ) \Big ).
\end{equation}
Making  use of this together with (\ref{HTa}), (\ref{HTb}) and Stirling's formula we deduce
\begin{eqnarray}\label{t1}
\lefteqn{
{e^{-x^2/2} H_{N-1}(x) \over 2^{N-1} \pi^{1/2} (N-2)!}
\int_0^{\sqrt{2N} X} e^{-t^2/2} H_{N-2}(t) \, dt } \nonumber \\
& & \sim  
{1 \over \pi \sqrt{2N} } {g_{-1,N}^{(H)}(X) \tilde{g}_{-2,N}^{(H)}(X) \over (1 - X^2) }
\Big ( 1 + O\Big ( {1 \over N} \Big ) \Big )\nonumber \\
 & & = 
{1 \over \pi \sqrt{2N} } {g_{-1,N}^{(H)}(X) \over (1 - X^2) } \Big (
X \tilde{g}_{-1,N}^{(H)}(X) + \sqrt{1 - X^2} g_{-1,N}^{(H)}(X) \Big )
\Big ( 1 + O\Big ( {1 \over N} \Big ) \Big )
\end{eqnarray}
where the equality follows from simple trigonometric identities.

For the second term in (\ref{H1'}), use of (\ref{H3}) shows
\begin{equation}\label{t2}
- {2^{-(N-1)} e^{-x^2} \over \pi^{1/2} (N-1)! }
( H_{N-1}(x))^2 \Big |_{x = \sqrt{2N} X} \: \sim \:
- {1 \over \pi} \sqrt{2 \over N}
{(g_{-1,N}^{(H)}(X))^2 \over \sqrt{1 - X^2} } \Big ( 1 + O\Big ( {1 \over N} \Big ) \Big ).
\end{equation}
And for the first term in (\ref{H1'}) we know from (\ref{2.3}) that
\begin{eqnarray}\label{t3}
\lefteqn{\rho( \sqrt{2N} X;{\rm UE}_{N}(e^{-x^2})) }  \nonumber \\
&& \sim
{\sqrt{2N} \over \pi} \sqrt{1 - X^2} 
- \sqrt{{2 \over N}} {\cos ( 2N X \sqrt{1 - X^2} + 2 N {\rm Arcsin} \, X - N \pi ) 
\over 2 \pi ( 1 - X^2)
} +  O \Big (
{1 \over N^{3/2} } \Big ).
\end{eqnarray}
But
\begin{equation}\label{1.b}
\cos( 2N X \sqrt{1 - X^2} + 2 N {\rm Arcsin} \, X - N \pi ) =
\sqrt{1 - X^2} (2 (\tilde{g}_{-1,N}^{(H)}(X))^2 - 1) - 2 X \tilde{g}_{-1,N}^{(H)}(X)
{g}_{-1,N}^{(H)}(X).
\end{equation}
Substituting (\ref{1.b}) in (\ref{t3}), then
adding this, 
(\ref{t1}),  and (\ref{t2}), and recalling (\ref{3.ke})  gives (\ref{H1a}).
\hfill $\square$

\medskip
The result (\ref{H1a}) agrees with that 
computed by Kalisch and Braak 
in \cite{KB02} and is also consistent with (\ref{3}). We
remark that in \cite{KB02} the $O(1/N^2)$ term is also 
given, being equal to
\begin{equation}\label{kbb}
{3 + 4 X^2 \over 16 \pi (1 - X^2)^{5/2} N^2 } -
{\cos((2N-1) {\rm Arcsin} \, X + 2N X \sqrt{1 - X^2}) \over 8 \pi (1 - X^2)^{5/2} N^2 }.
\end{equation}
In principle the present method offers a systematic approach to all correction terms.
For this we need the explicit form of the higher order terms in (\ref{Hm}), and these can in
fact be calculated from the results in \cite{PR34}. However we have not pursued such
calculations. We remark too that a calculation of the non-oscillatory portion of
(\ref{kbb}) is undertaken in \cite{DJ90}; however the result obtained does not agree
with (\ref{kbb}).
 
We turn our attention now to the GSE. First we note that
\begin{equation}\label{4.0}
\rho(X;{\rm SE}_N(e^{-4N x^2})) = 2 \sqrt{N}
\rho(2 \sqrt{N} X ;{\rm SE}_N(e^{-x^2})).
\end{equation}
Regarding the right hand side,
making use of (\ref{7.48a}),  (\ref{G1}) as well as the integral evaluation (see e.g.~\cite{AFNV00})
\begin{equation}\label{4.13'}
2^{-N} \int_0^\infty e^{-t^2/2} H_N(t) \, dt = \sqrt{\pi \over 2}
{N! \over 2^N (N/2)! }
\end{equation}
gives
\begin{eqnarray}\label{4.1}
\lefteqn{
\rho(x;{\rm SE}_N(e^{-x^2}))  
= {1 \over 2} \rho(x;{\rm UE}_{2N}(e^{-x^2}))  } \nonumber \\&& -
{e^{-x^2/2} H_{2N}(x) \over 4 \pi^{1/2} (2N-1)!}
\bigg ( {\sqrt{\pi \over 2} }
{(2N-1)! \over 2^{2N-1} (N-1/2)!} - 2^{-(2N-1)}
\int_0^x e^{-t^2/2} H_{2N-1}(t) \, dt \bigg ).
\end{eqnarray}
The asymptotic form of (\ref{4.1}) can be calculated according to the strategy of the proof of
Proposition \ref{p1a} to give the following result for the bulk scaled density in the GSE.

\begin{prop}
Let $-1 < X < 1$ be fixed, and $g_{0,2N}^{(H)}(X)$ be given according to (\ref{3.11}). We have
\begin{eqnarray}\label{tk}
{1 \over N} \rho(X;{\rm SE}_N(e^{-4Nx^2})) 
& \sim & 
{2 \over \pi} \sqrt{1 - X^2} - \Big ( {1 \over \sqrt{2 \pi N} } +  {(-1)^N \over 2 \pi N}
 \Big ) { g_{0,2N}^{(H)}(X) \over (1 - X^2)^{1/4} } \nonumber \\
&& \qquad + {1 \over 4 \pi N} {1 \over \sqrt{1 - X^2} }  +
O\Big ( {1 \over N^{3/2}} \Big ).
\end{eqnarray}
\end{prop}

\noindent
Proof. \quad Analogous to (\ref{HTa}), it follows from (\ref{Hm}) that
$$
\int_0^{2 \sqrt{N} X} e^{-t^2/2} H_{2N-1}(t) \, dt \: \sim \:
2 \sqrt{N}
\Big ( {2^{N-1} \over \pi^{1/4} N^{3/4} } ((2N)!)^{1/2} \Big )
\int_0^X {g_{-1,2N}^{(H)}(t) \over (1 - t^2)^{1/4} } \, dt,
$$
while proceeding as in the derivation of (\ref{HTb}) shows
$$
\int_0^X {g_{-1,2N}^{(H)}(t) \over (1 - t^2)^{1/4} } \, dt \: \sim \:
{1 \over 4 N} {1 \over (1 - X^2)^{3/4} }
\tilde{g}_{-1,2N}^{(H)} (X) - {(-1)^N \over 4 N}.
$$
Thus, after making use too of Stirling's formula,
\begin{eqnarray*}
\lefteqn{
 \Big ( {\sqrt{\pi \over 2} }
{(2N-1)! \over 2^{2N-1} (N-1/2)!} - 2^{-(2N-1)}
\int_0^{\sqrt{2N} X}  e^{-t^2/2} H_{2N-1}(t) \, dt \Big )
} \nonumber \\
&& \sim (N - 1/2)! \Big ( {1 \over \sqrt{2N} } - {2 \over \sqrt{\pi} } \Big (
{1 \over 4 N} {1 \over (1 - X^2)^{3/4} } \tilde{g}_{-1,2N}^{(H)}(X) - {(-1)^N \over 4 N}
\Big ) \Big ).
\end{eqnarray*}
Since (\ref{Hm}) gives
$$
e^{-x^2/2} H_{2N}(x) \Big |_{x = 2 \sqrt{N} X} = \pi^{-1/4}
{2^N N^{-1/4} \over (1 - X^2)^{1/4} } ((2N)!)^{1/2} \Big ( g_{0,2N}^{(H)}(X)
 + O\Big ( {1 \over N} \Big ) \Big )
$$
we thus have that with $x = 2 \sqrt{N} X$ the final line in (\ref{4.1}) has the asymptotic
behavior
$$
- {1 \over 4 \sqrt{\pi} } 
{g_{0,2N}^{(H)}(X) \over (1 - X^2)^{1/4} } \Big ( \sqrt{2} - {1 \over \sqrt{\pi N} }
\Big ( {1 \over (1 - X^2)^{3/4} } \Big ( X \tilde{g}_{0,2N}^{(H)}(X) +
\sqrt{1 - X^2} g_{0,2N}^{(H)}(X) \Big ) - (-1)^N \Big ).
$$
Further, we see from (\ref{2.3}) and (\ref{3.11}) that
\begin{eqnarray*}
\lefteqn{
{1 \over 2} \rho(2\sqrt{N} X ;{\rm UE}_{2N}(e^{-x^2}))  }
\\&&
\sim \:
{\sqrt{N} \over \pi} \sqrt{1-X^2} -
{\sqrt{1 - X^2} (2 (g_{0,2N}^{(H)}(X))^2 - 1) + 2 X \tilde{g}_{0,2N}^{(H)}(X) g_{0,2N}^{(H)}(X)
\over 8 \pi \sqrt{N}(1-X^2)} +
O \Big ( {1 \over N^{3/2} } \Big ).
\end{eqnarray*}
Adding these last two results, and recalling (\ref{4.1}) and (\ref{4.0})
gives the stated formula. \hfill $\square$

\medskip
In \cite{KB02}, at $O(1/N)$ only the non-oscillatory term
is reported. We note too that the non-oscillatory term at $O(1/N)$ in
(\ref{tk}) is consistent with (\ref{1.3'}) in the case $\beta = 4$.

\subsection{The LOE and LSE in the bulk}
We now apply the same strategy to the Laguerre case. For the LOE, substituting the appropriate
formula from (\ref{7.48a}), and (\ref{L1}), into (\ref{7.4.sum1}) shows
\begin{eqnarray}\label{aL1}
&& \rho(x;{\rm OE}_N(x^{(\alpha-1)/2} e^{-x/2}))  = 
\rho(x;{\rm UE}_{N-1}(x^{\alpha} e^{-x})) + {(N-1)! \over 4 (\alpha +N-2)!} 
x^{(\alpha -1)/2} e^{-x/2}
L_{N-1}^\alpha(x) \nonumber \\
&& \qquad \times \Big ( \int_0^\infty L_{N-2}^\alpha(t) t^{(\alpha-1)/2} e^{-t/2} \, dt -
2 \int_0^x  L_{N-2}^\alpha(t) t^{(\alpha-1)/2} e^{-t/2} \, dt \Big )
\end{eqnarray}
while
\begin{equation}\label{aL1a}
\rho(x;{\rm UE}_{N-1}(x^{\alpha} e^{-x})) =
\rho(x;{\rm UE}_{N}(x^{\alpha} e^{-x})) - {(N-1)! \over \Gamma(N + \alpha) }
\Big ( x^{\alpha/2} e^{-x/2} L_{N-1}^\alpha(x) \Big )^2.
\end{equation}

\begin{prop}\label{pL1}
Let $0 < X < 1$. We have
\begin{equation}\label{Lb1}
{1 \over N}  \rho( X; {\rm OE}_N(x^{(\alpha-1)/2} e^{-2Nx}) )
\: \sim \: \rho_{\rm MP}(X)  - {1 \over 2 \pi N}  {1 - \alpha \over \sqrt{ X (1 - X) } }
+ o(N^{-1}). 
\end{equation}
\end{prop}

\noindent
Proof. \quad 
By a change of variables
\begin{equation}\label{an}
\rho(X;{\rm OE}_N(x^{(\alpha - 1)/2} e^{-2Nx})) =
4 N \rho(4NX; {\rm OE}_N(x^{(\alpha - 1)/2} e^{-x/2})),
\end{equation}
so the task is to analyze the $N \to \infty$ asymptotics of the right hand
side of (\ref{aL1}) with $x$ replaced by $4NX$.

We know from \cite{NF95, AFNV00} that
\begin{equation}\label{4.21'}
\int_0^\infty L_{N-2}^a(t) t^{(a-1)/2} e^{-t/2} \, dt =
{\Gamma((N+1)/2) \Gamma(a+N-1) \over 2^{a/2 - 3/2} \Gamma(N) \Gamma((a+N)/2) }
\: \sim \: 2 N^{(a-1)/2}.
\end{equation}
Regarding the second integral in (\ref{aL1}), we first write
$$
\int_0^{4NX}  L_{N-2}^a(t) t^{(a-1)/2} e^{-t/2} \, dt =
\Big ( \int_0^{4 \epsilon} + \int_{4 \epsilon}^{4NX} \Big )
 L_{N-2}^a(t) t^{(a-1)/2} e^{-t/2} \, dt 
$$
where $0 < \epsilon \ll 1$.
In relation to the region $[4 \epsilon, 4NX]$, (\ref{Lm})  tells us that 
for $N$ even
$$
 L_{N-2}^\alpha(t) t^{a/2} e^{-t/2}  \Big |_{t \mapsto 4 N T} =
(2 \pi \sqrt{T(1-T)} )^{-1/2} N^{(\alpha-1)/2} \Big (
g_{-2,N}^{(L)}(T) + O\Big ( {1 \over N} \Big ) \Big ).
$$
Substituting in the integral and changing variables $T = s^2$ shows
\begin{equation}\label{4.19b}
\int_{4 \epsilon}^{4NX} L_{N-2}^\alpha(t) t^{(\alpha-1)/2} e^{-t/2} \, dt \: \sim \:
(4N)^{1/2} N^{(\alpha-1)/2} \Big ( {2 \over \pi} \Big )^{1/2} 
\int_{\sqrt{\epsilon/N}}^{\sqrt{X}} {1 \over \sqrt{s} (1 - s^2)^{1/4} }
g_{-2,N}^{(L)}(s^2) \, ds.
\end{equation}
In relation to the interval $t \in [0,4 \epsilon]$, we know that for $N \to \infty$
\cite{Sz75}
$$
L_{N-2}^\alpha(t) t^{\alpha/2} e^{-t/2} \: \sim \: N^{\alpha /2} J_{\alpha}(2(Nt)^{1/2}),
$$
where $J_n(z)$ denotes the Bessel function, and thus
\begin{equation}\label{4.19c}
 \int_0^{4 \epsilon} L_{N-2}^a(t) t^{(a-1)/2} e^{-t/2} \, dt \: \sim \:
N^{a/2} \int_0^{4 \epsilon} J_\alpha(2(Nt)^{1/2}) \, {dt \over \sqrt{t} }.
\end{equation}

We expect the leading contributions to come from the neighborhood of the upper terminal
$s = \sqrt{X}$ in (\ref{4.19b}), and the lower terminal $t=0$ in (\ref{4.19c})
(the integrands should connect smoothly from the lower terminal of (\ref{4.19b}) to the
upper terminal of (\ref{4.19c})). 
Since about $s = \sqrt{X}$
$$
g_{-2,N}^{(L)}(s^2) \: \sim \: \sin\Big (2N (\sqrt{X(1-X)} - {\rm Arcos} \, 
\sqrt{X}) - (\alpha -3)
{\rm Arcos} \, \sqrt{X} + 3 \pi/4 + 4N \sqrt{1-X} (s - \sqrt{X}) \Big )
$$
we have
$$
\int^{\sqrt{X}} {g_{-2,N}^{(L)}(s^2) \over \sqrt{s} (1 - s^2)^{1/4} } \, ds
\: \sim \:
- {1 \over 4 N \sqrt{1 - X} } 
{ \tilde{g}_{-2,N}^{(L)}(X) \over  X^{1/4} (1 - X)^{1/4} } 
$$
where
$$
\tilde{g}^{(L)}_{m,N}(x) := \cos \Big ( 2N ( \sqrt{x(1-x)} - 
{\rm Arcos} \, \sqrt{x} ) -
(\alpha+1+2m) {\rm Arcos} \, \sqrt{x} + 3 \pi / 4 \Big ).
$$
Also
$$
\int_0^\infty J_\alpha (t) \, dt = 1,
$$
so we have
$$
\int_0  J_\alpha (2(Nt)^{1/2}) \, {dt \over \sqrt{t} } \: \sim \: N^{-1/2}.
$$
Reading off the asymptotic form of the factor 
$x^{a/2} e^{-x/2} L_{N-1}^a(x)|_{x=4N X}$
in (\ref{aL1}) from (\ref{Lm}) we deduce from this that
$$
\rho(4NX;{\rm OE}_N(x^{(\alpha -1)/2} e^{-x/2})) \:  \sim  \:
\rho(4NX;{\rm UE}_{N-1} (x^{\alpha} e^{-x}))   -
{ g_{-1,N}^{(L)}(X) \tilde{g}_{-2,N}^{(L)}(X) \over 8 \pi N X (1 - X) } 
$$

It remains to determine the asymptotic form of (\ref{aL1a}). For this we use 
the analogue of (\ref{an}), and (\ref{Lm}), to obtain 
$$
\rho(4NX;{\rm UE}_{N-1} (x^{\alpha} e^{-x}))  \: \sim \:
{1 \over 4 N} \rho(X;{\rm UE}_{N} (x^{\alpha} e^{-2 N x}))
 - {1 \over 2 \pi} \sqrt{ {1 \over X(1-X)} } (g_{-1,N}(X))^2.
$$
Noting from the definitions and by a simple trigonometric identity that
\begin{eqnarray*}
\tilde{g}^{(L)}_{-2,N}(X) & = & (2 X - 1) \tilde{g}^{(L)}_{-1,N}(X) -
2 \sqrt{(1-X)X} {g}^{(L)}_{-1,N}(X) \\
\tilde{g}^{(L)}_{0,N}(X) & = & (2 X - 1) \tilde{g}^{(L)}_{-1,N}(X) +
2 \sqrt{(1-X)X} {g}^{(L)}_{-1,N}(X) 
\end{eqnarray*}
we therefore have
\begin{equation}\label{sf1}
\rho(4NX;{\rm OE}_N(x^{(\alpha-1)/2} e^{-x/2})) \: \sim \:
{1 \over 4 N} \rho (X; {\rm UE}_N(x^\alpha e^{-4Nx})) - {1 \over 8 \pi N}
{g^{(L)}_{-1,N}(X) \tilde{g}_{0,N}^{(L)}(X) \over X (1 - X) }.
\end{equation}
Now, with
\begin{equation}\label{4.25'}
A_{N,\alpha}(X) := 2 N( \sqrt{X(1-X)} - {\rm Arccos} \, \sqrt{X} ) - \alpha {\rm Arccos} \, \sqrt{X}
\end{equation}
we have that
\begin{eqnarray*}
g^{(L)}_{-1,N}(X) & = & \sin ( A_{N,\alpha}(X) + {\rm Arc cos } \, \sqrt{X} + 3 \pi / 4) \\
\tilde{g}^{(L)}_{0,N}(X) & = & \cos ( A_{N,\alpha}(X) - {\rm Arc cos } \, \sqrt{X} + 3 \pi / 4)
\end{eqnarray*}
and thus
\begin{equation}\label{sf2}
g^{(L)}_{-1,N}(X) \tilde{g}^{(L)}_{0,N}(X) = {1 \over 2} \Big ( - \cos (2 A_{N,\alpha}(X)) +
2 \sqrt{X(1-X)} \Big ).
\end{equation}
Substituting (\ref{sf2}) in (\ref{sf1}) and noting from (\ref{2.7}) that
\begin{equation}\label{sf3}
{1 \over 4 N} \rho(X;{\rm UE}_N(x^\alpha e^{-4 N x}) ) \: \sim \:
{1 \over 4} \rho_{\rm MP}(X) - {\cos 2 A_{N,\alpha}(X) \over 16 \pi X (1-X) N} 
+ {\alpha \over 8 \pi \sqrt{X(1-X)} N }
\end{equation}
we obtain (\ref{Lb1}).
\hfill $\square$

\medskip
Consider now the LSE. Analogous to (\ref{an}), by a change of variables
$$
\rho(X;{\rm SE}_N(x^{\alpha+1} e^{-8Nx})) = 8 N \rho ( 8N X; {\rm SE}_N(x^{\alpha+1} e^{-x})),
$$
while (\ref{7.nfs}) together with the fact \cite{NF95,AFNV00} 
\begin{equation}\label{4.26'}
\int_0^\infty e^{-t/2} t^{(\alpha-1)/2} L_{2N-1}^\alpha(t) \,dt = 0
\end{equation}
shows that
\begin{eqnarray}\label{w.1}
\lefteqn{
\rho(x;{\rm SE}_N(x^{\alpha+1} e^{-x})) =
{1 \over 2} \rho(x;{\rm UE}_{2N}(x^\alpha e^{-x})) } \nonumber \\
&& \qquad -
{\Gamma(1+2N) \over 4 \Gamma(\alpha+2N) }
e^{-x/2} x^{(\alpha-1)/2} L_{2N}^\alpha(x)
\int_0^x e^{-t/2} t^{(\alpha-1)/2} L_{2N-1}^\alpha(t) \,dt.
\end{eqnarray}

\begin{prop}
Let $0 < X < 1$. In terms of the notation (\ref{gL}) and (\ref{4.25'}) we have
\begin{eqnarray}\label{Lb4}
\lefteqn{
{1 \over N}  \rho( X; {\rm SE}_N(x^{\alpha+1} e^{-8Nx}) )
\: \sim \: \rho_{\rm MP}(X)  } \nonumber
\\ &&
- {1 \over 2(\pi N)^{1/2}} {g_{0,2N}^{(L)}(X) \over X^{3/4} (1 - X)^{1/4}}
+
{\alpha + 1 \over 4 \pi N \sqrt{X(1-X)} }
+ o(N^{-1}).
\end{eqnarray}
\end{prop}

\noindent
Proof. \quad 
Following the strategy of the proof of Proposition \ref{pL1}
we find
$$
\int_0^{8NX}  e^{-t/2} t^{(\alpha-1)/2} L_{2N-1}^\alpha(t) \,dt \: \sim \:
(2N)^{(\alpha - 1)/2} +
{(2N)^{\alpha/2} \over (2 \pi)^{1/2} }
{\tilde{g}_{-1,2N}^{(L)}(X) \over 2N X^{1/4}(1 - X)^{3/4}}.
$$
Also, (\ref{Lm}) and Stirling's formula show
$$
 {\Gamma(1+2N) \over 4 \Gamma(\alpha+2N) }
e^{-x/2} x^{(\alpha-1)/2} L_{2N}^\alpha(x) \Big |_{x = 8NX} \: \sim \:
 {1 \over 16 (\pi N)^{1/2} (2N)^{(\alpha-1)/2}}
{g_{0,2N}^{(L)}(X) \over X^{3/4} (1 - X)^{1/4} }.
$$
Substituting in (\ref{w.1}) gives
\begin{eqnarray}\label{sf4}
 \rho(8NX;{\rm SE}_N(x^{\alpha+1} e^{-x})) & \sim & {1 \over 16 N}
\rho(X;{\rm UE}_{2N}(x^\alpha e^{-8Nx}))
\nonumber \\&&
- {1 \over 16 (\pi N)^{1/2} } {g_{0,2N}^{(L)}(X) \over X^{3/4} (1 - X)^{1/4}}
- {g_{0,2N}^{(L)}(X) \tilde{g}_{-1,2N}^{(L)}(X) \over 32 \pi N X (1 - X) }.
\end{eqnarray}
Analogous to (\ref{sf2}) we have
$$
g_{0,2N}^{(L)}(X)  \tilde{g}_{-1,2N}^{(L)}(X) = - {1 \over 2} \Big (
\cos 2 A_{2N,\alpha}(X) + 2 \sqrt{X(1-X)} \Big ).
$$
Substituting this together with (\ref{sf3}) in (\ref{sf4}) gives (\ref{Lb4}).
\hfill $\square$

\section{Asymptotic expansions at the edge}
\setcounter{equation}{0}
\subsection{The GOE and GSE}
The scaled densities $\rho(X;{\rm OE}_N(e^{-NX^2}))$ and $\rho(X;{\rm SE}_N(e^{-4NX^2}))$ have
to leading order
their support on $(-1,1)$. We know from previous studies \cite{Fo93,FNH99} that setting $X$
as specified by (\ref{XZ}) (with the restriction $\xi < 0 $ removed) and multiplying by $N^{1/3}$,
the limit $N \to \infty$ exists and can be computed explicitly. Here we are interested
in computing the first correction, as in the soft edge formula (\ref{GUE edge result}) for the GUE.

In the case of the GOE, we see from (\ref{3.ke}), (\ref{H1}) and (\ref{H1'}) that in addition to
the knowledge of (\ref{GUE edge result}), an asymptotic formula for
$\rho(1 + \xi/2 N^{2/3};{\rm OE}_N(e^{-Nx^2}))$ can be obtained by making use of (\ref{xt1}).

\begin{prop}
We have
\begin{eqnarray}\label{G1s}
&& {1 \over 2 N^{2/3} } \rho \Big ( 1 + {\xi \over 2 N^{2/3} } ; {\rm OE}_N(e^{-N x^2}) \Big )
\nonumber \\
&  & \qquad = (\aip(\xi))^2 - \xi (\ai(\xi))^2 + {1 \over 2} \ai(\xi) \Big ( 1 - 
\int_\xi^\infty \ai(t) \, dt \Big ) \nonumber \\
&  & \qquad \quad + {1 \over 2 N^{1/3} } \aip(\xi) \Big ( 1 - 
\int_\xi^\infty \ai(t) \, dt \Big ) + O \Big ( {1 \over N^{2/3} } \Big ).
\end{eqnarray}
\end{prop}

\noindent
Proof. \quad Consider the integral in (\ref{H1}). We know from \cite{AFNV00} that
$$
\int_0^x e^{-t^2/2} H_N(t) \, dt = \sqrt{ \pi \over 2} {N! \over (N/2)!} -
\int_x^\infty e^{-t^2/2} H_N(t) \, dt.
$$
Replacing $N$ by $N-2$, setting $x = (2N)^{1/2} + 2^{-1/2} N^{-1/6} \xi$, making use
of (\ref{xt1}) and simplifying shows that
\begin{eqnarray}\label{m3}
&&
\int_0^{(2N)^{1/2} + 2^{-1/2} N^{-1/6} \xi} e^{-t^2/2} H_{N-2}(t) \, dt \nonumber \\
&& \quad = \sqrt{\pi \over 2} {(N-2)! \over (N/2 - 1)!} \bigg ( 1 -
\int_\xi^\infty \ai(y) \, dy + {2 \over N^{1/3} } \ai(\xi) +
O\Big ( {1 \over N^{2/3} } \Big ) \bigg ).
\end{eqnarray}
Now using (\ref{xt1}) with $m=-1$, and multiplying with the result (\ref{m3}) as required by
(\ref{H1}) we obtain
\begin{eqnarray*}
&&
\bigg ({e^{-x^2/2} H_{N-1}(x) \over 2^{N-1} \pi^{1/2} (N-2)! }
\int_0^x e^{-t^2/2} H_{N-2}(t) \, dt \bigg ) \Big |_{x = (2N)^{1/2} +
2^{-1/2} N^{-1/6} \xi} \nonumber \\
&& \qquad 
\sim {N^{1/6} \over 2^{1/2} } \Big ( \ai(\xi) + {1 \over N^{1/3} } \aip(\xi) \Big )
\Big ( 1 - \int_\xi^\infty \ai(y) \, dy + {2 \over N^{1/3} } \ai(\xi) \Big ).
\end{eqnarray*}

According to (\ref{H1'}), we also require the asymptotic formula
$$
- {2^{-N+1} \over \pi^{1/2} (N-1)!} ( e^{-x^2/2} H_{N-1}(x) )^2
\Big |_{x = (2N)^{1/2} + 2^{-1/2} N^{-1/6} \xi} \: \sim \:
- {\sqrt{2} \over N^{1/6} } (\aip(\xi) )^2,
$$
which follows from (\ref{xt1}). Further
$$
\rho((2N)^{1/2} + 2^{-1/2} N^{-1/6} \xi; {\rm UE}_N(e^{-x^2}) ) =
{1 \over (2N)^{1/2} } \rho(1 + \xi/2 N^{2/3} ; {\rm UE}_N (e^{- 2 N x^2} ) )
$$
(cf.~(\ref{3.ke})) so the asymptotic form of the first term on the the right
hand side of (\ref{H1'}) can be read off from (\ref{GUE edge result}). Doing this, and
recalling (\ref{3.ke}), we deduce (\ref{G1s}). \hfill $\square$

\medskip
We turn our attention now to the GSE.
We will analyze (\ref{4.1}) rewritten to read
\begin{equation}\label{xtp}
\rho(x;{\rm SE}_N(e^{-x^2}) ) =
{1 \over 2} \rho(x;{\rm UE}_{2N}(e^{-x^2})) -
{e^{-x^2/2} H_{2N}(x) \over 2^{2N+1} \pi^{1/2} (2N-1)!}
\int_x^\infty e^{-t^2/2} H_{2N-1}(t) \, dt.
\end{equation}

\begin{prop}
We have
\begin{eqnarray}\label{ix}
&&
{1 \over (2N)^{2/3} } \rho \Big ( 1 + {\xi \over 2 (2N)^{2/3} }; {\rm SE}_N(e^{-4Nx^2}) \Big )
\nonumber \\
&&
\quad \sim \: (\aip(\xi))^2 - \xi (\ai(\xi))^2 - {1 \over 2} \ai(\xi) \bigg (
\int_\xi^\infty \ai(t) \, dt - {1 \over (2N)^{1/3} } \ai(\xi) + O(N^{-2/3}) \bigg ).
\nonumber \\
\end{eqnarray}
\end{prop}

\noindent
Proof. \quad By a simple change of variables
\begin{equation}\label{xtp1}
\rho \Big ( 1 + {\xi \over 2(2N)^{2/3} }; {\rm SE}_N(e^{- 4 N x^2} ) \Big ) =
2 \sqrt{N} \rho \Big (
(4N)^{1/2} + 2^{-1/2} (2N)^{-1/6} \xi; {\rm SE}_{N}(e^{-x^2}) \Big ),
\end{equation}
so we seek the large $N$ asymptotic form of (\ref{xt1}) with
$x \mapsto (4N)^{1/2} + 2^{-1/2} (2N)^{-1/6} \xi$. Making use of (\ref{xt1}) shows
\begin{eqnarray}\label{xtp2}
&&
\bigg ({e^{-x^2/2} H_{2N}(x) \over 2^{2N+1} \pi^{1/2} (2N-1)!}
\int_x^\infty e^{-t^2/2} H_{2N-1}(t) \, dt \bigg )
\Big |_{x = (4N)^{1/2} + 2^{-1/2} (2N)^{-1/6} \xi} \nonumber \\
&& \qquad \sim
{N^{1/6} \over 2^{4/3} }  \ai(\xi) \Big \{
\int_\xi^\infty \ai(t) \, dt - {1 \over (2N)^{1/3} } \ai(X) + O(N^{-2/3})
\Big \},
\end{eqnarray}
while it follows from (\ref{GUE edge result}) that
\begin{equation}\label{xtp3}
{1 \over 2} \rho\Big ((4N)^{1/2} + 2^{-1/2} (2N)^{-1/6} \xi ; {\rm UE}_{2N}(e^{-x^2}) \Big )
\: \sim \: {(2N)^{1/6} \over \sqrt{2} } \bigg (
(\aip(\xi))^2 - \xi (\ai(\xi))^2 + O\Big ( {1 \over N^{2/3} } \Big ) \bigg ).
\end{equation}
Substituting (\ref{xtp3}) and (\ref{xtp2}) in (\ref{xtp}) and recalling
(\ref{xtp1}) gives the stated result.
\hfill $\square$

\subsection{The LOE and LSE}
The soft edge scaling variables for the LOE and LSE are the same as those for the LUE,
exhibited in (\ref{LUE edge result}). The leading correction term to the corresponding
soft edge densities are given by the following result.

\begin{prop}\label{edge prop}
We have
\begin{eqnarray}\label{op1}
&&{1 \over (2N)^{2/3}} \rho \Big ( 1 + {\xi \over  (2N)^{2/3} } ; {\rm OE}_N( x^{(\alpha - 1)/2}
e^{-2Nx}) \Big ) \nonumber \\
&& \quad
\: \sim \: 
(\aip(\xi))^2 - \xi (\ai(\xi))^2 + {1 \over 2} \ai(\xi) 
\Big ( 1 -
\int_\xi^\infty \ai(s) \, ds \Big ) \nonumber \\
&& \qquad
- {(\alpha - 1 ) \over  2 (2N)^{1/3}} \Big [ \aip(\xi) \Big (
1 - \int_\xi^\infty \ai(s) \, ds \Big ) -  (\ai(\xi))^2 \Big ]
+ O(N^{-2/3}) 
\end{eqnarray}
and
\begin{eqnarray}\label{op2}
&&{2 \over (4N)^{2/3}} \rho \Big ( 1 + {\xi \over  (4N)^{2/3} } ; {\rm SE}_N( x^{\alpha + 1}
e^{-8Nx}) \Big ) \: \sim \: 
(\aip(\xi))^2 - \xi (\ai(\xi))^2 - {1 \over 2} \ai(\xi) 
\int_\xi^\infty \ai(s) \, ds  \nonumber \\
&& \qquad + {\alpha + 1 \over  2 (4N)^{1/3}} \Big ( (\ai(\xi))^2  +
\aip(\xi) \int_\xi^\infty \ai(t) \, dt \Big )
+ O(N^{-2/3}).
\end{eqnarray}
\end{prop}

\noindent
Proof. \quad Consider first the LOE. We rewrite (\ref{aL1}) as
\begin{eqnarray}\label{aL1'}
&& \rho(x;{\rm OE}_N(x^{(\alpha-1)/2} e^{-x/2}))  = 
\rho(x;{\rm UE}_{N-1}(x^\alpha e^{-x}) +
{(N-1)! \over 4 (\alpha + N - 2)! } x^{(\alpha-1)/2} e^{-x/2} L_{N-1}^\alpha(x) \nonumber \\
&& \qquad \times \Big (
2 \int_x^\infty L_{N-2}^\alpha(t) t^{(\alpha-1)/2} e^{-t/2} \, dt -
\int_0^\infty L_{N-2}^\alpha(t) t^{(\alpha-1)/2} e^{-t/2} \, dt \Big ).
\end{eqnarray}
The asymptotic form of the final integral in (\ref{op2}) is known from (\ref{4.21'}).
According to (\ref{an}),  we seek the asymptotic form of the remaining terms with
$$
x = 4N +  2 (2N)^{1/3} \xi.
$$
For the first integral, use of (\ref{3.17}) gives
\begin{eqnarray*}
&&
\int_{4N+2(2N)^{1/3}\xi}^\infty L_{N-2}^\alpha(t) t^{(\alpha - 1)/2} e^{- t/2} \, dt
\\ && \qquad
\: \sim \: N^{(\alpha - 1)/2} \Big (
\int_\xi^\infty \ai(s) \, ds - {3 - \alpha
\over (2 N)^{1/3} } \ai(\xi) + O(N^{-2/3}) \Big ).
\end{eqnarray*}
The asymptotic form of the term outside the bracketed integrals is given by (\ref{3.17}) with
$p=-1$. For the first term on the right hand side of (\ref{aL1'}), use of (\ref{aL1a}),
the analogue of (\ref{an}), and (\ref{LUE edge result}) shows
\begin{eqnarray*}
&&
\rho(4N +  2(2N)^{1/3}\xi;{\rm UE}_{N-1}(x^\alpha e^{-x}))  \nonumber \\
&& \quad \sim {1 \over 2 (2N)^{1/3} } \Big ( (\aip(\xi))^2 - \xi (\ai(\xi))^2 +
{\alpha \over 2^{1/3} } (\ai(\xi))^2 {1 \over N^{1/3}} \Big ) 
 - 2^{-2/3} N^{-2/3} (\ai(\xi))^2.
\end{eqnarray*}
The asymptotic form of all terms have now been determined, and (\ref{op1}) follows.

In relation to the LSE, we use (\ref{4.26'}) to rewrite (\ref{w.1}) to read 
\begin{eqnarray}\label{w.1A}
\lefteqn{
\rho(x;{\rm SE}_N(x^{\alpha+1} e^{-x})) =
{1 \over 2} \rho(x;{\rm UE}_{2N}(x^\alpha e^{-x})) } \nonumber \\
&& \qquad +
{\Gamma(1+2N) \over 4 \Gamma(\alpha+2N) }
e^{-x/2} x^{(\alpha-1)/2} L_{2N}^\alpha(x)
\int_x^\infty e^{-t/2} t^{(\alpha-1)/2} L_{2N-1}^\alpha(t) \,dt.
\end{eqnarray}
A simple change of variables gives
\begin{equation}\label{US}
\rho \Big ( 1 + {\xi \over (4N)^{2/3} }; {\rm SE}_{N}(x^{\alpha+1} e^{-8Nx}) \Big ) =
8 N \rho \Big ( 8N  + 2(4N)^{1/3} \xi; {\rm SE}_{N}(x^{\alpha+1} e^{-x})
\Big ),
\end{equation}
so we seek the asymptotic form of (\ref{w.1A}) with
$$
x =  8N +  2(4N)^{1/3} \xi.
$$
For this, we make use of (\ref{3.17}) to deduce
\begin{eqnarray*}
\lefteqn{
{\Gamma(1+2N) \over 4 \Gamma(\alpha+2N) }
e^{-x/2} x^{(\alpha-1)/2} L_{2N}^\alpha(x)
\int_x^\infty e^{-t/2} t^{(\alpha-1)/2} L_{2N-1}^\alpha(t) \,dt }
 \nonumber \\
&& \sim - {1 \over 8} 2^{-2/3} N^{-1/3} \Big ( \ai(\xi) 
- {\alpha + 1 \over (4N)^{1/3} } \aip(\xi) \Big )
\Big (
\int_\xi^\infty \ai(s) \, ds + {\alpha - 1
\over (4N)^{1/3} }\ai(\xi) + O(N^{-2/3}) \Big ).
\end{eqnarray*}
Further, making use of the analogue of (\ref{US}) and recalling (\ref{LUE edge result}) shows
\begin{eqnarray*}
\lefteqn{
{1 \over 2}
\rho  (8N + 2(4N)^{1/3} \xi; {\rm UE}_{2N}(x^{\alpha} e^{-x}) ) } \nonumber \\
&& \sim {1 \over 4 (4N)^{1/3} } \Big (
(\aip(\xi))^2 - \xi (\ai(\xi))^2 + {\alpha \over 2^{1/3} (2N)^{1/3} }(\ai(\xi))^2 
+ O(N^{-2/3})  \Big ).
\end{eqnarray*}
The asymptotic form of all terms in (\ref{w.1A}) are now determined. After use of (\ref{US}),
(\ref{op2}) follows.
\hfill  $\square$ 

Figure 1 provides a numerical comparison of the
asymptotic expansion given by (\ref{op2}) with the
exact result for the LSE density given by (\ref{w.1A}) and (\ref{US}).

\begin{figure}
\includegraphics{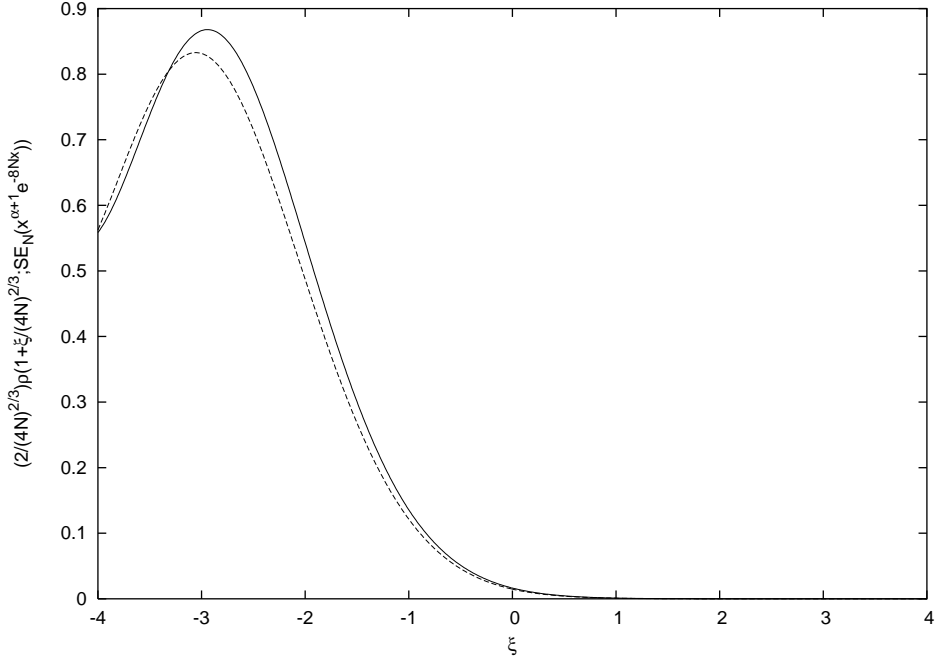}
\caption{ Numerical comparison of the
asymptotic expansion (\ref{op2}), shown as the dashed line, and the
exact result for the LSE density given by (\ref{w.1A}) and (\ref{US}),
shown as the solid line, for the eigenvalue density near
the soft edge of the LSE with $N=20$ and $\alpha=1/2$.}
\end{figure}

\section{Consequences}
\subsection{Matching the bulk and edge expansions}
\setcounter{equation}{0}
\subsection*{The GOE and GSE}
In Section \ref{s2.2} the conjectured matching between the bulk asymptotic expansion expanded
about the soft edge, and the asymptotic expansion of the edge density expanded into the bulk,
for both the GUE and LUE was given further credence by its confirmation on further terms in the
asymptotic expansion. To study the matching in the case of orthogonal and symplectic
symmetry, it is therefore convenient to express the corresponding asymptotic expansions in
terms of the unitary symmetry expansions. We will consider first the Gaussian cases.

According to (\ref{H1a}), (\ref{kbb}), (\ref{2.3}), (\ref{2.5a})
\begin{eqnarray}\label{k1}
&& {1 \over N} \rho(X;{\rm OE}_N(e^{-Nx^2})) \: \sim \:
{1 \over N} \rho(X;{\rm UE}_N(e^{-Nx^2})) +
\Big ( {\cos[ 2 N \pi P_{\rm W}(X)] \over 4 \pi (1 - X^2) } -
{1 \over 2 \pi \sqrt{1 - X^2} } \Big ) {1 \over N} \nonumber \\
&& \qquad + \Big ( {1 + 2 X^2 \over 8 \pi (1 - X^2)^{5/2} } -
{X(21 + 2 X^2) \sin[2N\pi P_{\rm W}(X)] \over 48 \pi (1 - X^2)^{5/2} } -
{\cos[2 N \pi P_{\rm W}(X)] \over 8 \pi (1 - X^2)^2 } \Big ) {1 \over N^2};
\end{eqnarray}
according to (\ref{G1s}) and (\ref{GUE edge result})
\begin{eqnarray}\label{k2}
&& {1 \over 2N^{2/3}} \rho \Big ( 1 + {\xi \over 2 N^{2/3} } ; {\rm OE}_N(e^{-Nx^2}) \Big )
 \: \sim \:
 {1 \over 2N^{2/3}} \rho \Big ( 1 + {\xi \over 2 N^{2/3} } ; {\rm UE}_N(e^{-2Nx^2}) \Big )
\nonumber \\
&& \quad + {1 \over 2} \ai(\xi) \Big ( 1 - \int_\xi^\infty \ai(t) \, dt \Big )
+ {1 \over 2 N^{1/3}} \aip(\xi)  \Big ( 1 - \int_\xi^\infty \ai(t) \, dt \Big ) +
O \Big ( {1 \over N^{2/3}} \Big );
\end{eqnarray}
according to (\ref{tk}) and (\ref{2.3})
\begin{eqnarray}\label{k3}
&& {1 \over N} \rho(X;{\rm SE}_N(e^{-4Nx^2})) \: \sim \: 
{1 \over 2N} \rho(X;{\rm UE}_{2N}(e^{-2Nx^2})) +
{\cos[ 4N \pi P_{\rm W}(X)] \over 8 \pi (1 - X^2) N}
\nonumber \\
&& \quad 
{1 \over 4 \pi N \sqrt{1 - X^2} }  -
\Big ( {1 \over \sqrt{2 \pi N}} + {1 \over 2 \pi N} \Big )
{\cos[ 2N \pi P_{\rm W}(X) + {1 \over 2} {\rm Arcsin} \, X ] \over  (1 - X^2)^{1/4}} +
O(N^{-3/2});
\end{eqnarray}
according to (\ref{ix}) and (\ref{GUE edge result})
\begin{eqnarray}\label{k4}
&& {1 \over (2N)^{2/3}} \rho \Big ( 1 + {\xi \over 2 (2N)^{2/3} } ; {\rm SE}_N(e^{-4Nx^2}) \Big )
 \: \sim \:
 {1 \over 2(2N)^{2/3}} \rho \Big ( 1 + {\xi \over 2 (2N)^{2/3} } ; {\rm UE}_{2N}(e^{-4Nx^2}) \Big )
\nonumber \\
&&
- {1 \over 2} \ai(\xi) \int_\xi^\infty \ai(t) \, dt  
 + {(\ai(\xi))^2 \over 2 (2N)^{1/3}} 
+ O \Big ( {1 \over N^{2/3}} \Big );
\end{eqnarray}

Substituting (\ref{XZ}) in (\ref{k1}) and (\ref{k3}) and expanding as in (\ref{tm}) gives
\begin{eqnarray}\label{c1}
{1 \over 2 N^{2/3}} \rho\Big ( 1 + {\xi \over 2 N^{2/3} } ; {\rm OE}_N(e^{-N x^2}) \Big )
& \mathop{\sim}\limits^\cdot &
{1 \over 2 N^{2/3}} \rho\Big ( 1 + {\xi \over 2 N^{2/3} } ; {\rm UE}_N(e^{-2N x^2}) \Big )
\nonumber \\
&& + {3 \over 16 \pi |\xi|^{5/2} }  +
{23 \sin(4|\xi|^{3/2}/3) \over 96 \pi |\xi|^{5/2} } 
 + {\cos(4 |\xi|^{3/2}/3) \over 8 \pi |\xi| }
\nonumber \\
&& - \Big ( {1 \over 4 \pi |\xi|^{1/2} } + {\cos(4 |\xi|^{3/2}/3) \over 16 \pi |\xi|^2}
\Big ) {1 \over N^{1/3} },
\end{eqnarray}
and
\begin{eqnarray}\label{c2}
&& {1 \over (2N)^{2/3}} \rho\Big ( 1 + {\xi \over 2 (2N)^{2/3} } ; {\rm SE}_N(e^{-4 N x^2}) \Big )
 \mathop{\sim}\limits^\cdot 
{1 \over 2 (2N)^{2/3}} \rho\Big ( 1 + {\xi \over 2 (2N)^{2/3} } ; {\rm UE}_{2N}(e^{-4N x^2}) \Big )
\nonumber \\
&& \qquad - {1 \over 2} {1 \over \sqrt{\pi} } {\sin( {2 \over 3} |\xi|^{3/2} + \pi/4) \over 
|\xi|^{1/4} } +
{\cos( {4 \over 3} |\xi|^{3/2}) \over 8 \pi |\xi| } \nonumber \\
&& \qquad + \Big ( {1 \over 4 \pi |\xi|^{1/2} } +
{|\xi|^{1/4} \sin( {2 \over 3} |\xi|^{3/2} - \pi/4) \over 4 \sqrt{\pi} } \Big )
{1 \over (2N)^{1/3} } + O(N^{-1/2}). 
\end{eqnarray}

We now compare (\ref{c1}) and (\ref{c2}) to the $\xi \to -\infty$ expansion of
(\ref{k2}) and (\ref{k4}) respectively. In relation to the latter, a straightforward
calculation using the leading two terms of (\ref{aia}) shows that for $|\xi| \to \infty$,
$$
\int_{|\xi|}^\infty \ai(-t) \, dt \: \sim \: {1 \over \sqrt{\pi} } \Big (
{1 \over |\xi|^{3/4} } \cos( {2 \over 3} |\xi|^{3/2} + \pi/4) + {41 \over 48}
{1 \over |\xi|^{9/4}} \sin( {2 \over 3} |\xi|^{3/2} + \pi/4) \Big ).
$$ 
Using this, together with (\ref{aia}) itself, we find that (\ref{c1}) agrees
with the $\xi \to - \infty$ expansion
of (\ref{k2}) for the first two terms of the $O(1)$ part in (\ref{c1}), but the 
rational factor of ${1 \over 8}$ should be ${1 \over 4}$ for the third term. At $O(N^{-1/3})$
agreement is obtained with the first term.
Similarly, we find that the the first term at each order in (\ref{c2}) agrees with the
$\xi \to -\infty$ expansion of (\ref{c2}). These results are all consistent with the
matching hypothesis. 

\subsection*{The LOE and LSE}
As in the Gaussian cases, we begin by expressing the densities for the LOE and LSE in terms
of the corresponding densities for the LUE.

According to (\ref{Lb1}) and (\ref{2.7})
\begin{eqnarray}\label{6.7}
&&
{1 \over N} \rho(X;{\rm OE}_N(x^{(\alpha -1)/2} e^{-2Nx})) 
\sim {1 \over N} \rho(X;{\rm UE}_N(x^{\alpha } e^{-4Nx})) 
 \nonumber \\
&&   \qquad +
\Big ( {\cos2 A_{N,\alpha}(X)  \over
 4 \pi X (1 - X) } - {1 \over 2 \pi (X(1-X))^{1/2} } \Big ) {1 \over N};
\end{eqnarray}
according to (\ref{op1}) and (\ref{LUE edge result})
\begin{eqnarray}\label{6.8}
&&
{1 \over (2N)^{2/3}} \rho(1 + {\xi \over (2N)^{2/3}};{\rm OE}_N(x^{(\alpha -1)/2} e^{-2Nx})) 
\: \sim \: {1 \over (2N)^{2/3}} \rho(1 + {\xi \over (2N)^{2/3}};{\rm UE}_N(x^{\alpha } e^{-4Nx})) 
 \nonumber \\
&& \qquad + {1 \over 2} \ai(\xi) \Big ( 1 - \int_\xi^\infty \ai(s) \, ds \Big ) -
{(\alpha - 1) \over 2 (2N)^{1/3} }
\aip(\xi) \Big ( 1 - \int_\xi^\infty \ai(s) \, ds \Big ) 
- {(\alpha + 1) \over 2 (2N)^{1/3} } (\ai(\xi))^2
;
\end{eqnarray}
according to (\ref{Lb4}) and (\ref{2.7})
\begin{eqnarray}\label{6.9}
&&
{1 \over N} \rho(X;{\rm SE}_N(x^{\alpha +1} e^{-8Nx})) 
\: \sim \:  {1 \over 2 N} \rho(X;{\rm UE}_{2N}(x^{\alpha } e^{-8Nx}))
 \nonumber \\
&& \qquad - {1 \over 2 (\pi N)^{1/2}} {g_{0,2N}^{(L)}(X) \over X^{3/4} (1 - X)^{1/4}}
+ {\cos 2 A_{2N,\alpha}(X)  \over 8 \pi N X (1 - X) } +
{1 \over 4 \pi N \sqrt{X(1-X)} };
\end{eqnarray}
according to (\ref{op2}) and (\ref{LUE edge result})
\begin{eqnarray}\label{6.10}
&&
{2 \over (4N)^{2/3}} \rho(1 + {\xi \over (4N)^{2/3}};{\rm SE}_N(x^{\alpha +1} e^{-8Nx}))
\: \sim \: {1 \over (4N)^{2/3}} \rho(1 + {\xi \over (4N)^{2/3}};{\rm UE}_{2N}(x^{\alpha } e^{-8Nx}))
 \nonumber \\
&&\qquad - {1 \over 2} \ai(\xi) \int_\xi^\infty \ai(s) \, ds  +
{(\alpha + 1) \over 2 (4N)^{1/3} } \aip(\xi) \int_\xi^\infty \ai(t) \, dt -
{(\alpha - 1) \over 2 (4N)^{1/3} }
(\aip(\xi))^2 .
\end{eqnarray}

Substituting (\ref{XZ}) in (\ref{6.7}) and (\ref{6.9}) and expanding gives
\begin{eqnarray}\label{6.13}
&&
{1 \over (2N)^{2/3}} \rho(1 + {\xi \over (2N)^{2/3}};{\rm OE}_N(x^{(\alpha -1)/2} e^{-2Nx}))
\: \mathop{\sim}\limits^\cdot \: {1 \over (2N)^{2/3}} \rho(1 + 
{\xi \over (2N)^{2/3}};{\rm UE}_N(x^{\alpha } e^{-4Nx}))
 \nonumber \\
&& \qquad + {\cos(4 |\xi|^{3/2}/3) \over 4 \pi |\xi|} -
{(1 + \alpha \sin(4 |\xi|^{3/2}/3)) \over 2 \pi \sqrt{|\xi|} } {1 \over (2N)^{1/3} }
\end{eqnarray}
and
\begin{eqnarray}\label{6.14}
&&
{2 \over (4N)^{2/3}} \rho(1 + {\xi \over (4N)^{2/3}};{\rm SE}_N(x^{\alpha +1} e^{-8Nx}))
\:  \mathop{\sim}\limits^\cdot 
\: {1 \over (4N)^{2/3}} \rho(1 + {\xi \over (4N)^{2/3}};{\rm UE}_{2N}(x^{\alpha } e^{-8Nx}))
 \nonumber \\
&&\qquad + {\sin(2|\xi|^{3/2}/3-3\pi/4) \over 2 \sqrt{\pi} |\xi|^{1/4} }   +
{ \cos(4|\xi|^{3/2}/3) \over 4 \pi |\xi|} \nonumber \\
&& \qquad
+ {1 \over 2 (4N)^{1/3} \pi \sqrt{|\xi|} }
\Big ( 1 +  (1 + \alpha) \sqrt{\pi} |\xi|^{3/4} \cos(2 |\xi|^{3/2} - 3 \pi/4) 
-
 {\alpha \over 2^{1/3} } \sin(4|\xi|^{3/2}/3) \Big ).
\end{eqnarray}

On the other hand let us expand (\ref{6.8}) and (\ref{6.10}) for $\xi \to - \infty$.
Doing so we find agreement with  the first term at each order in (\ref{6.13}) and
(\ref{6.14}) respectively, as consistent with the matching hypothesis.

\subsection{Microscopic delta functions}
The results of Sections 4 and 5 tell us the asymptotic expansion of the global
density, and the soft edge density. Here we would like to relate these
expansions to the result (\ref{1.3'}) and its Laguerre analogue.

Consider first the Gaussian cases. For $|\xi|$ large but otherwise
arbitrary, we write
\begin{eqnarray}\label{sat}
&& \int_{-\infty}^\infty \rho(x;{\rm ME}_N(e^{-\beta x^2/2})) a(x) \,
dx  \nonumber \\
&&
= \Big ( \int_{R_1} + \int_{R_2} \Big ) 
\rho\Big ( X; {\rm ME}_N(e^{-N \beta x^2})) \tilde{a}(X) \, dX 
 \nonumber \\
&& =
\int_{R_1} \rho(X;{\rm ME}_N(e^{-N \beta x^2})) \tilde{a}(X) \, dX
 \nonumber \\
&&
+ {1 \over 2 N^{2/3} } \Big (
\int_{-|\xi|}^\infty \rho(1+ y/2N^{2/3}; {\rm ME}_N (e^{-\beta N x^2})) 
\tilde{a}(1 + y/2N^{2/3}) \, dy
\nonumber \\
&&
+ 
\int^{|\xi|}_{-\infty}  \rho(1- y/2N^{2/3}; {\rm ME}_N (e^{-\beta N x^2}))
\tilde{a}(-1 - y/2N^{2/3}) \, dy \Big )
\end{eqnarray}
where ME${}_N $ $=$ ${\rm OE}_N$, ${\rm UE}_N$, ${\rm SE}_N$ respectively,
$R_1 = (-1 + |\xi|/2N^{2/3}, 1 - |\xi|/2N^{2/3})$ and
$R_2 = (-\infty,\infty)\backslash R_1$, 
$\tilde{a}(x) = a(Nx)$
and we have used the fact that $\rho$ is
even. Because
\begin{equation}\label{rsoft}
\rho^{\rm soft}(y;{\rm ME}_N(e^{-\beta N x^2})) :=
\lim_{N \to \infty} {1 \over 2 N^{2/3}}
\rho(1 + y/2N^{2/3};{\rm ME}_N(e^{-\beta N x^2}))
\end{equation}
is an $O(1)$ quantity, we see that to leading order the second and third
integrals in (\ref{sat}) contribute
\begin{equation}\label{sat1}
(\tilde{a}(1) + \tilde{a}(-1))
\int_{-|\xi|}^\infty \rho^{\rm soft}(y;{\rm ME}_N(e^{-\beta N x^2})) \, dy.
\end{equation}
However, in relation to the first integral on the right hand side of
(\ref{sat}), we know that terms which are different order in $N$ in the
asymptotic expansion of $\rho(X;{\rm ME}_N(e^{-N\beta x^2}))$ can contribute
at the same order upon the substitution (\ref{XZ}). Unlike the situation at
the edge, the asymptotic expansion of this integral does not therefore
correspond directly to the
asymptotic of the integrand, leaving us without a method of analysis.
Nonetheless some insight into the microscopic origin of the delta functions
in (\ref{1.3'}) can be obtained as a consequence of the functional form of
(\ref{rsoft}) for $\beta = 1, 4$.

For $\beta = 2$ we read off from (\ref{GUE edge result})
$$
\rho^{\rm soft}(y;{\rm UE}_N(e^{-2 N x^2})) =
(\aip(y))^2 - y (\ai(y))^2,
$$
while (\ref{tm}) tells us that the leading $y \to -\infty$ behavior is
$2 \sqrt{|y|}/\pi$ so (\ref{sat1}) diverges for $|\xi| \to \infty$.
Because of the result (\ref{1.3'}) it must be that this is exactly canceled by a
contribution from the bulk, and thus the edge terms (\ref{sat1}) cancel.

For $\beta = 1$ and 4 we observe from (\ref{G1s}) and (\ref{ix}) that
$\rho^{\rm soft}(y;{\rm UE}_N(e^{-2Nx^2})$ appears as an additive component
in the scaled soft edge density, together with a further term in both cases.
The further term has the property that it is integrable for $y \to -\infty$.
Thinking then of the decomposition (\ref{sat}) for $|\xi|$ large, and ignoring
the contribution from the non-integrable additive component, this then suggests
that
\begin{eqnarray}\label{rs2a}
&&\int_{-\infty}^\infty \rho(x;{\rm OE}_N(e^{-x^2/2})) a(x) \, dx  \sim  
N \int_{-1}^1 \rho_{\rm W}(X) \tilde{a}(X) \, dX - {1 \over 2 \pi}
\int_{-1}^1 {\tilde{a}(X) \over \sqrt{1 - X^2} } \, dX
\nonumber \\
&& + (\tilde{a}(1) + \tilde{a}(-1)) {1 \over 2}
\int_{-\infty}^\infty \ai(y) \Big ( 1 - \int_y^\infty \ai(t) \, dt \Big ) dy
\end{eqnarray}
\begin{eqnarray}\label{rs2b}
&&\int_{-\infty}^\infty \rho(x;{\rm SE}_N(e^{-2 x^2})) a(x) \, dx  \sim 
N \int_{-1}^1 \rho_{\rm W}(X) \tilde{a}(X) \, dX + {1 \over 4 \pi}
\int_{-1}^1 {\tilde{a}(X) \over \sqrt{1 - X^2} } \, dX
\nonumber \\
&& - (\tilde{a}(1) + \tilde{a}(-1)) {1 \over 4}
\int_{-\infty}^\infty \ai(y) \Big (  \int_y^\infty \ai(t) \, dt \Big ) dy.
\end{eqnarray}
Here the bulk contributions are the leading two non-oscillatory terms exhibited in
(\ref{H1a}) and (\ref{tk}) respectively, and the edge contributions are the
leading terms in (\ref{G1s}) and (\ref{ix}) respectively, with the component
corresponding to the $\beta = 2$ edge deleted. Since
\begin{eqnarray}
\ai(y) \Big ( 1 - \int_y^\infty \ai(t) \, dt \Big ) & = &
{1 \over 2} {d \over dy} \Big ( 1 - \int_y^\infty \ai(t) \, dt \Big )^2 \\
\ai(y)   \int_y^\infty \ai(t) \, dt  & = &
- {1 \over 2} {d \over dy} \Big ( \int_y^\infty \ai(t) \, dt \Big )^2
\end{eqnarray}
we see from $\int_{-\infty}^\infty \ai(t) \, dt = 1$ that the final integrals
in (\ref{rs2a}) and (\ref{rs2b}) are both equal to $1/2$, thus reclaiming
(\ref{1.3'}). 

We consider now the Laguerre analogue of (\ref{1.3'}). Let us introduce the
so called chiral matrix ensembles chME${}_N(g(x))$ by the eigenvalue p.d.f.
\begin{equation}\label{chE}
{1 \over C} \prod_{l=1}^N g(x_l) x_l^{\beta/2} \prod_{1 \le j < k \le N}
|x_k^2 - x_j^2|^\beta, \qquad (x_l > 0).
\end{equation}
The simple change of variables $x_j^2 \mapsto x_j$ shows
\begin{equation}\label{acc1}
{1 \over 2} \rho(X;{\rm chME}_N(x^{2(a+(2-\beta)/4)}
e^{-2N\beta x^2})) =
X \rho(X^2;{\rm ME}_N(x^a e^{-2 N \beta x})),
\end{equation}
and thus that the Laguerre ensembles can be viewed as a Gaussian version
of the chiral ensemble, generalized by the factor
$x^{2(a+(2-\beta)/4)}$. We see from (\ref{acc1}), (\ref{Lb1}), (\ref{2.7}) 
and (\ref{Lb4}) that
\begin{eqnarray*}
&& \rho(X;{\rm chOE}_N(x^{a/2} e^{-2Nx^2})) \: \sim \:
2 \rho_{\rm W}(X) + {1 \over \pi N}
{a - 1/2 \over \sqrt{1 - X^2} } \\
&& \rho(X;{\rm chUE}_N(x^{a} e^{-4Nx^2})) \: \sim \:
2 \rho_{\rm W}(X) + 
{a  \over \pi N \sqrt{1 - X^2} } \\
&&  \rho(X;{\rm chSE}_N(x^{2a} e^{-8Nx^2})) \: \sim \:
2 \rho_{\rm W}(X) + 
{a  + 1/4 \over \pi N \sqrt{1 - X^2} }
\end{eqnarray*}
where only non-oscillatory terms are included. In the case $a=0$ these
expansions are precisely the same as for the corresponding Gaussian ensembles
(the leading term in the chiral ensembles is $2 \rho_{\rm W}(X)$ rather than
$\rho_{\rm W}(X)$ because $X \in (0,1)$ rather than $(-1,1)$).

At the soft edge $X = 1$ of the chiral ensembles, the fact that the scaled
densities are the same as for the Gaussian ensembles tells us that there is
a contribution
$$
{1 \over 2 N} \Big ( {1 \over \beta} - {1 \over 2} \Big )
\delta (X - 1)
$$
to the smoothed density. And, although not the focus of attention of the
present work, for the Laguerre and thus chiral ensembles there is an
edge effect at $X=0$ (the so called hard edge \cite{Fo93}) which one expects
to give a microscopic contribution
$$
- {a \over 2N} \delta^+(X)
$$
(see the next subsection)
where $\int_0^\infty f(X) \delta^+(X) \, dX = f(0)$.
Consequently we expect the Laguerre analogue of (\ref{1.3'}) to be
\begin{eqnarray}\label{1.4'}
\lefteqn{
\rho(X;{\rm chME}_N(x^{\beta a/2} e^{-2 \beta N x^2})) } \nonumber \\
&& \sim
2 \rho_{\rm W}(X) + {a \over \pi N \sqrt{1 - X^2}} -
{a \over 2 N} \delta^+(X) + {1 \over N}
\Big ( {1 \over \beta} - {1 \over 2} \Big ) \Big (
\delta(X-1) - {1 \over \pi} {1 \over \sqrt{1 - X^2}} \Big ).
\end{eqnarray}

\subsection{Macroscopic balance}
In this final subsection, we will show that the results
(\ref{1.3'}), (\ref{1.4'}) are consistent with macroscopic considerations.
 
In a one-component log-gas,
to leading order in $N$ the electrostatic potential created by the particle density
$\sigma(x)$ must balance the background potential $u(x)$, and thus the equation
\begin{equation}\label{uA}
u(x) + C = \int_{-c}^c \sigma(y) \log |x - y| \, dy, \qquad
x \in (-c,c)
\end{equation}
where $C$ is {\it some} constant, must hold. This is to be solved subject to the
particle conservation constraint
\begin{equation}\label{uAa}
\int_{-c}^c \sigma(y) \, dy = 1.
\end{equation}
For example, with $u(x) = x^2$, (\ref{uA}) and (\ref{uAa}) are satisfied with
\begin{equation}
\sigma(y) = \rho_{\rm W}(y)
\end{equation}
(see e.g.~\cite{Fo02}). To the next order in $N$, fluctuations in the particle
density create an electric field and thus a force density which must be
balanced for the system to be stable. The balancing force is provided by the
gradient of the pressure fluctuation, and leads to the refinement of
(\ref{uA}) \cite{Dy62'}
\begin{equation}\label{uAp}
u(x) + C = \int_{-c}^c \sigma(y) \log |x - y| \, dy
+ {1 \over N} \Big ( {1 \over 2} - {1 \over \beta} \Big ) \log \sigma (x), \qquad
x \in (-c,c)
\end{equation}
which again is subject to (\ref{uAa}). With $u(x) = x^2$, setting
\begin{equation}
\sigma(y) = \rho_{\rm W}(y) + {\mu(y) \over N}
\end{equation}
we see that
\begin{equation}\label{uAc}
C = \int_{-c}^c \mu(y) \log |x - y| \, dy +
\Big ( {1 \over 2} - {1 \over \beta} \Big ) \log \rho_{\rm W}(x), \qquad
x \in (-c,c)
\end{equation}
which must be solved subject to the constraint
\begin{equation}\label{uAd}
\int_{-c}^c \mu(y) \, dy = 0.
\end{equation} 
Differentiating (\ref{uAc}) gives
\begin{equation}\label{uAe}
0 = {\rm PV} \int_{-1}^1 {\mu(y) \over x - y} \, dy -
\Big ( {1 \over 2} - {1 \over \beta} \Big )
{x \over 1 - x^2},
\end{equation}
where PV denotes the principal value. 
Making use of the fact that
$$
{\rm PV} \int_{-1}^1 {1 \over x - y}
{1 \over \sqrt{1 - y^2} } \, dy = 0, \qquad x \in (-1,1)
$$
(see e.g.~\cite{PS90}), we see that (\ref{uAe}), (\ref{uAd}) is solved by
$$
\mu(y) = \Big ( {1 \over \beta} - {1 \over 2} \Big )
\Big ( {1 \over 2} \Big ( \delta(y-1) + \delta(y+1) \Big ) -
{1 \over \pi} {1 \over \sqrt{1 - y^2} } \Big )
$$
in agreement with (\ref{1.3'}).

The chiral ensemble (\ref{chE}) is a log-gas confined to $x>0$ with image
charges in $x <0$. At leading order in $N$ balance of the electric field
requires
\begin{equation}\label{kA}
u(x) + C = \int_0^c \sigma(y) \log | x^2 - y^2| \, dy, \qquad
x \in (-c,c),
\end{equation}
subject to the constraint
$$
\int_0^c \sigma(y) \, dy = 0.
$$
But (\ref{kA}) can be written
$$
u(x) + C = \int_{-c}^c \sigma(y) \log |x - y| \, dy
$$
so we have essentially the previous situation. 
In particular, with $u(x) = 2x^2$, this shows (\ref{kA}) is satisfied with
\begin{equation}
\sigma(y) = 2 \rho_{\rm W}(y).
\end{equation}
To next order in $N$ the chiral ensembles in (\ref{1.4'}) have
$u(x) = 2x^2 - {a \over 2N} \log x$, and the generalization of (\ref{uAc})
reads
$$
- {a \over 2} \log x + C =
\int_{-1}^1 \mu(y) \log |x - y| \, dy +
\Big ( {1 \over 2} - {1 \over \beta} \Big ) \log \rho_{\rm W}(x).
$$
With $\mu(y) \mapsto \mu(y) - {a \over 2} \delta(x)$ this is in fact
identical to
(\ref{uAc}) and we thus reclaim (\ref{1.4'}).


\begin{thebibliography}{10}

\bibitem{AFNV00}
M.~Adler, P.J. Forrester, T.~Nagao, and P.~van Moerbeke,
\newblock ``Classical skew orthogonal polynomials and random matrices,''
\newblock {\em J. Stat. Phys.} {\bf 99}, 141\---170 (2000).

\bibitem{DJ90}
G.S. Dhesi and R.C. Jones.
\newblock ``Asymptotic corrections to the Wigner semicircular eigenvalue spectrum
  of a large real symmetric random matrix using the replica method,''
\newblock {\em J. Phys. A} {\bf 23}, 5577\---5599 (1990).

\bibitem{Dy62'}
F.J. Dyson.
\newblock ``Statistical theory of energy levels of complex systems {II},''
\newblock {\em J. Math. Phys.} {\bf 3}, 157\---165 (1962).

\bibitem{Po65}
C.E.~Porter (editor).
\newblock {\em Statistical theories of spectra: fluctuations}.
\newblock Academic Press, New York, 1965.

\bibitem{Fo02}
P.J. Forrester.
\newblock Log-gases and {Random} {Matrices}.
\newblock www.ms.unimelb.edu.au/\~{}matpjf/matpjf.html.

\bibitem{Fo93}
P.J. Forrester,
\newblock ``The spectrum edge of random matrix ensembles,''
\newblock {\em Nucl. Phys. B} {\bf 402}, 709\---728 (1993).

\bibitem{FNH99}
P.J. Forrester, T.~Nagao, and G.~Honner,
\newblock ``Correlations for the orthogonal-unitary and symplectic-unitary
  transitions at the hard and soft edges,''
\newblock {\em Nucl. Phys. B} {\bf 553}, 601\---643 (1999).

\bibitem{GFF05}
T.M. Garoni, P.J. Forrester, and N.E. Frankel,
\newblock ``Asymptotic corrections to the eigenvalue density of the {GUE} and
  {LUE},''
\newblock arXiv:math-ph/0504053.

\bibitem{Jo98}
K.~Johansson.
\newblock ``On fluctuation of eigenvalues of random {Hermitian} matrices''
\newblock {\em Duke Math. J.} {\bf 91}, 151\---204 (1998).

\bibitem{KB02}
F.~Kalisch and D.~Braak,
\newblock ``Exact density of states for finite {G}aussian random matrix ensembles
  via supersymmetry,''
\newblock {\em J. Phys. A} {\bf 35}, 9957\---9969 (2002).

\bibitem{Moecklin34}
E.~Moecklin,
\newblock ``Asymptotische Entwicklungen der Laguerreschen Polynome,''
\newblock {\em Comm. Math. Helv.} {\bf 7}, 24\---46 (1934).


\bibitem{NF95}
T.~Nagao and P.J. Forrester,
\newblock ``Asymptotic correlations at the spectrum edge of random matrices,''
\newblock {\em Nucl. Phys. B} {\bf  435}, 401\---420 (1995).

\bibitem{Ol74}
F.~Olver.
\newblock {\em Asymptotics and Special Functions}.
\newblock Academic Press, London, 1974.

\bibitem{PR34}
M.~Plancherel and W.~Rotach,
\newblock ``Sur les valeurs asymptotiques des polynomes d'{H}ermite,''
\newblock {\em Comm. Math. Helv.} {\bf 1}, 227\---254 (1929).

\bibitem{PS90}
D.~Porter and D.S.G. Stirling.
\newblock {\em Integral equations: a practical treatment, from spectral theory
  to applications}.
\newblock Cambridge University Press, 1990.

\bibitem{Sz75}
G.~Szeg\"o.
\newblock {\em Orthogonal polynomials}.
\newblock American Mathematical Society, Providence R.I., 4th edition, 1975.

\bibitem{VZ84} 
J.J.M. Verbaarschot and M.R. Zirnbauer,
\newblock{``Replica variables, loop expansion, and spectral rigidity of random-matrix ensembles,''}
\newblock {\em Ann. Phys.} {\bf 158}, 78\---119 (1984).

\end{thebibliography}

\end{document}